\documentclass[twocolappendix, twocolumn]{aastex701}

\usepackage{comment}
\usepackage{tabularx}
\usepackage{booktabs}
\usepackage{amsmath}
\usepackage{graphicx}
\usepackage{capt-of}

\newcolumntype{Y}{>{\centering\arraybackslash}X}

\newcommand{\argmax}{\mathop{\rm argmax}\limits} 

\newcommand{\asiaa}{Academia Sinica Institute of Astronomy and Astrophysics, 11F of ASMA Building, No.1, Sec. 4, Roosevelt Rd, Taipei 106319, Taiwan}

\newcommand{\kyushu}{Department of Earth and Planetary Sciences, Faculty of Sciences, Kyushu University, Nishi-ku, Fukuoka 819-0395, Japan}

\newcommand{\naoj}{National Astronomical Observatory of Japan, National Institutes of Natural Sciences, 2-21-1 Osawa, Mitaka, Tokyo 181-8588, Japan}

\newcommand{\nsyu}{Department of Physics, National Sun Yat-Sen University, No. 70, Lien-Hai Road, Kaohsiung City 80424, Taiwan, R.O.C.}
\newcommand{\cag}{Center of Astronomy and Gravitation, National Taiwan Normal University, Taipei 116, Taiwan}

\shortauthors{Yamaguchi et al.}
\received{March 21, 2026}
\revised{May 29, 2026}
\accepted{June 22, 2026}
\journalinfo{Accepted for publication in The Astrophysical Journal}
\begin{document}
\title{A Hybrid Origin for the Multiple Ring–Gap Structures in the Large Protoplanetary Disk V1094 Sco: A Low-Mass Planet and Secular Gravitational Instability}


\correspondingauthor{Masayuki Yamaguchi}
\author[orcid=0000-0002-8185-9882,gname=Masayuki,sname=Yamaguchi]{Masayuki Yamaguchi}
\affil{\kyushu}
\affil{\naoj}
\email[show]{yamaguchi.masayuki.376@m.kyushu-u.ac.jp}

\author[orcid=0000-0002-0963-0872,gname=Masahiro, sname=N. Machida]{Masahiro N. Machida}
\affiliation{\kyushu}
\email{machida.masahiro.018@m.kyushu-u.ac.jp}

\author[0000-0002-8596-3505]{Ryosuke T. Tominaga}
\affiliation{Department of Earth and Planetary Sciences, Institute of Science Tokyo, 2-12-1, Ookayama, Meguro, Tokyo 152-8551, Japan}
\email{tominaga.r.aa@m.titech.ac.jp}

\author[0000-0003-4361-5577]{Jinshi Sai}
\affiliation{Department of Physics and Astronomy, Graduate School of Science and Engineering, Kagoshima University, 1-21-35 Korimoto, Kagoshima, Kagoshima 890-0065, Japan}
\email{jn.insa.sai@gmail.com}
\email{jinshi.sai@sci.kagoshima-u.ac.jp}

\author{Takayuki Muto}
\affiliation{Division of Liberal Arts, Kogakuin University, 1-24-2 Nishi-Shinjuku, Shinjuku, Tokyo 163-8677, Japan}
\email{muto@cc.kogakuin.ac.jp}

\author[0000-0001-9248-7546]{Michihiro Takami}
\affil{\asiaa}
\email{hiro@asiaa.sinica.edu.tw}

\author[0000-0003-2300-2626]{Hauyu Baobab Liu}
\affil{\nsyu}
\affil{\cag}
\email{baobabyoo@gmail.com}

\author[0000-0001-6580-6038]{Ayumu Shoshi}
\affiliation{Department of Earth and Planetary Sciences, Graduate School of Science, Kyushu University, 744 Motooka, Nishi-ku, Fukuoka 819-0395, Japan}
\email{shoshi.ayumu.660@s.kyushu-u.ac.jp}

\author[0000-0002-6034-2892]{Takashi Tsukagoshi}
\affiliation{Faculty of Engineering, Ashikaga University, Ohmae-cho 268-1, Ashikaga, Tochigi 326-8558, Japan}
\email{takashi.tsukagoshi.astro@gmail.com}

\author[0009-0001-1889-9216]{Shu Ishibashi}
\affiliation{Department of Physics and Astronomy, Graduate School of Science and Engineering, Kagoshima University, 1-21-35 Korimoto, Kagoshima, Kagoshima 890-0065, Japan}
\email{k8174036@kadai.jp}


\begin{abstract}
High spatial resolution observations reveal that some protoplanetary disks host multiple ring-gap pairs at large stellocentric radii, yet their physical origin remains unsettled. We present a multiwavelength analysis of the V1094~Sco disk using Atacama Large Millimeter/submillimeter Array Band~6 continuum and $^{12}$CO and $^{13}$CO $J=2-1$ emission, together with a Very Large Telescope/SPHERE near-infrared scattered-light image. The continuum image shows four narrow dust ring-gap pairs extending to exceptionally large radii ($r \sim 380$ au), while the CO isotopologues trace a spatially extended gas disk ($r \sim 760$ au) in Keplerian rotation. From the dust ring widths, we place conservative upper limits on the turbulent viscosity parameter, $\alpha \lesssim 10^{-3}$ and potentially $\lesssim 10^{-4}$, implying weak turbulence. The ensemble of gap widths and depths is inconsistent with a simple one-planet-per-gap interpretation. At $r \simeq 100$~au, a double gap and its scattered-light counterpart are consistent with multigap excitation by a single low-mass companion of $(55 \pm 35)\,M_{\oplus}$. At $r \simeq 170$-$230$~au, the outer ring system shows regular spacing and no clear scattered-light counterpart, indicating mechanisms that operate primarily at the disk midplane. These outer rings are quantitatively compatible with secular gravitational instability. V1094~Sco therefore supports a hybrid pathway in which weak turbulence in an extended disk allows secular gravitational instability to assemble long-lived midplane dust concentrations that can cradle planet formation beyond $\sim100$~au, alongside planet-driven substructures at intermediate radii.
\end{abstract}




\section{Introduction} \label{sec:intro}

Planets form within protoplanetary disks through the dynamical and collisional evolution of gas and solids \citep{Hayashi1985}. Over the past decade, high spatial resolution observations, particularly with the Atacama Large Millimeter/submillimeter Array (ALMA), have revealed that disk substructures such as rings and gaps are nearly ubiquitous in Class~II disks \citep[e.g.,][]{yen_gas_2016, tsukagoshi_gap_2016, tang_planet_2017, hashimoto_asymmetric_2021, ueda_massive_2022, orihara_alma_2023, liu_first_2024, shoshi2025b}. While many of these features are now known to appear as early as the Class~I stage \citep{shoshi2025a}, their physical origin remains debated. Planet--disk interactions provide a compelling explanation in many systems \citep{yamaguchi_alma_2024}, but multiple alternative mechanisms---including snowline-related dust evolution \citep[e.g.,][]{okuzumi_sintering-induced_2016, pinilla_dust_2017}, magnetically driven processes \citep[e.g.,][]{suriano_formation_2019, hu_nonideal_2019, riols_ring_2020}, and hydrodynamic or gravitational instabilities \citep[e.g.,][]{youdin_formation_2011, takahashi_origin_2016, tominaga_secular_2023}---can also produce qualitatively similar morphologies. Disentangling these scenarios remains one of the central challenges in connecting disk substructures to planet formation.

This challenge becomes particularly acute when substructures appear as multiple ring--gap pairs at large stellocentric radii. In such disks, a straightforward one--planet--per--gap interpretation can quickly become problematic, both in terms of formation timescales and long-term dynamical stability \citep{ndugu_are_2019, tzouvanou_all_2023}. Progress therefore requires benchmark disks in which the radial locations, widths, and contrasts of multiple ring--gap pairs are robustly measured, enabling quantitative tests of whether the observed pattern is consistent with planet-driven gaps or instead favors alternative mechanisms \citep[e.g.,][]{yamaguchi_alma_2024}.

The protoplanetary disk around the T~Tauri star V1094~Sco provides an outstanding laboratory in this context. This object is located in the Lupus~3 star-forming region \citep{tachihara_13co_1996, Hara1999PASJ} at a distance of $154.76 \pm 0.76$~pc \citep[Gaia~DR3;][]{Gaia2023} and hosts one of the most massive and radially extended disks with multiple dust ring-gap pairs known among T~Tauri stars. The system was first highlighted by its unusually strong millimeter emission with Atacama Submillimeter Telescope Experiment (ASTE) and a cold temperature structure inferred from spectral energy distribution (SED) modeling \citep[][]{tsukagoshi_detection_2011}. Early ALMA observations at $\sim0\farcs3$ resolution resolved a pair of prominent ring--gap structures \citep{van_terwisga_v1094_2018} and more recent ALMA observations and analyses at $\sim0\farcs1$ resolution revealed increasingly complex radial oscillations in the dust continuum emission \citep{vioque_alma_2025}.

Despite these advances, the physical origin of the ring--gap architecture in V1094~Sco remains unclear. The coexistence of rich dust substructures, the cold outer disk, and a large dust reservoir raises fundamental questions about which mechanisms dominate at different radii and vertical layers, and how such structures relate to planet formation in the outer disk.

In this work, we present a multiwavelength analysis of V1094~Sco based on uniformly reprocessed ALMA Band~6 continuum and CO isotopologues data, together with an archival Very Large Telescope (VLT)/SPHERE near-infrared polarimetric image. By compiling all suitable ALMA executions and performing flux rescaling, astrometric alignment, and iterative self-calibration prior to imaging, we construct high-sensitivity visibility datasets and apply super-resolution imaging to achieve the highest spatial resolution of $0\farcs04$ currently available for this target. This enables us to resolve the ring--gap architecture in unprecedented detail. Section~\ref{sec:data_imaging} summarizes the datasets, calibration, self-calibration, and imaging products. Section~\ref{sec:analysis_results} reports the key observational results, including the continuum morphology and disk geometry, CO kinematics, disk radial extent, the hierarchy of continuum substructures, and their comparison with scattered-light. Section~\ref{sec:analysis} derives physical properties anchored to these observations, including the dynamical stellar mass, the surface geometry from scattered-light, and disk temperature. Section~\ref{sec:discussion} discusses the implications for weak turbulence and possible origins of the ring--gap architecture including planet--disk interactions and secular gravitational instability (hereafter secular GI). Section~\ref{sec:conclusion} summarizes the main findings.

\section{Data and Imaging}
\label{sec:data_imaging}

In this section, we summarize the observational data and imaging products used throughout this paper. We analyzed ALMA 12 m array continuum and CO isotopologues line datasets obtained in both compact and extended configurations. We compiled all suitable ALMA archival executions and combined them into a single visibility dataset to achieve high sensitivity while retaining high angular resolution. A summary of the datasets is given in Table~\ref{tab:alma_obs}. In addition, we used an archival near-infrared scattered-light dataset obtained with VLT/SPHERE in the $H$ band. Below, we describe the data products and the minimal processing steps for the ALMA and VLT/SPHERE images.

\subsection{ALMA Calibration and Self-calibration}
\label{sec:alma_calibration}

All ALMA datasets were calibrated in the Common Astronomy Software Applications package \citep[\texttt{CASA};][]{CASATeam2022}, starting from the pipeline calibrated measurement sets for each execution block. Because our analysis relied on multiepoch observations and multiple array configurations, we applied three steps prior to concatenation to suppress dominant systematics: (i) astrometric alignment of the phase centers across epochs, (ii) relative flux rescaling across epochs using overlapping $uv$ ranges, and (iii) iterative self-calibration on the continuum using our developed tool \texttt{AJISAI}. These steps were required to prevent artificial image blurring and spurious substructures when constructing high-fidelity combined images. The full technical procedure, including CASA task level details and table handling, is provided in Appendices~\ref{apx:clean_imaging_details} and \ref{apx:selfcal}.

\subsection{Continuum Imaging}
\label{sec:continuum_imaging}

Using the self-calibrated data, we adopted a dual image strategy for the continuum. This approach separates the analysis of high-resolution morphology from beam-convolved images used for noise characterization.

For the main morphological analyses, we used super-resolution imaging with $\tt PRIISM$\footnote[1]{$\tt PRIISM$ (Python Module for Radio Interferometry Imaging with Sparse Modeling) is a public ALMA imaging tool based on sparse modeling, available at \url{https://github.com/tnakazato/priism}.} \citep[version 0.11.5;][]{Nakazato2020, Nakazato_priism_2020}. PRIISM reconstructs a model image via regularized maximum-likelihood optimization with an $\ell_1$ term and a total squared variation (TSV) term, and selects regularization parameters via cross validation \citep[CV; ][]{yamaguchi_super-resolution_2020}. We used the CV selected model image as our primary high-resolution representation of the dust continuum emission. The model image has units of $\rm Jy~pixel^{-1}$ because this imaging process does not include beam convolution. The effective spatial resolution $\theta_{\rm eff}$ was measured with the point-source injection method \citep{yamaguchi_alma_2021} and achieved $\theta_{\rm eff} = 40 \times 30$~mas at a PA of $-77\arcdeg$. This represents a factor of two improvement compared to the standard CLEAN image made with Briggs robust $=0.5$ ($\theta_{\rm cl} = 97\times74$~mas at a PA of $-88\arcdeg.7$).

For analyses that require a beam-convolved image and an empirical rms noise estimate, we also used a PRIISM restored image constructed following \citet{Yamaguchi_alma_2025ApJ}. This image has units of $\rm Jy~beam^{-1}$, a synthesized beam of $362 \times 331$~mas (PA $=87\arcdeg$), and an rms noise level of $23~\mu\rm Jy~beam^{-1}$ measured in emission-free regions. We use this restored image primarily to characterize faint extended continuum emission and to provide a noise-referenced comparison with the high-resolution PRIISM model image.

A quantitative assessment of the PRIISM performance, including the imaging procedure, measurement of effective spatial resolution via point-source injection, validation in the visibility domain, and the restoration procedure, is provided in Appendix~\ref{apx:priism_image}.

\subsection{CO line imaging}
\label{sec:co_imaging}

The $^{12}$CO ($J=2\textrm{--}1$) and $^{13}$CO ($J=2\textrm{--}1$) line measurement sets were prepared following the calibration and data-combination workflow described in Appendix~\ref{apx:clean_imaging_details}. In brief, we applied the relative flux scaling derived from the continuum visibilities and the continuum self-calibration gain tables to the line data, and then combined the selected datasets prior to $uv$-plane continuum subtraction.

While the continuum image was reconstructed using $\tt PRIISM$ to improve the spatial resolution, the spectral line cubes were imaged with the conventional CLEAN algorithm for two main reasons. First, the signal-to-noise ratio (SNR) per velocity channel is substantially lower than in the continuum image, whereas the spatial resolution improvement achieved by $\tt PRIISM$ depends on image sensitivity, with higher SNR yielding better effective resolution \citep{yamaguchi_alma_2024}. Second, reconstructing a large number of velocity channels with $\tt PRIISM$ would be computationally prohibitive.

We therefore reconstructed the $^{12}$CO and $^{13}$CO cubes using the multiscale CLEAN algorithm, adopting imaging parameters that prioritize sensitivity to extended emission over spatial resolution. Specifically, we used (1) clean masks generated by the \texttt{automasking} algorithm, (2) Briggs robust $=2.0$ (comparable to natural weighting), and (3) a $uv$ taper with an FWHM of $500~\mathrm{k}\lambda$. The velocity channel width was $0.16~\mathrm{km~s^{-1}}$ per channel for $^{12}$CO and $0.17~\mathrm{km~s^{-1}}$ for $^{13}$CO. This setup resulted in synthesized beams of $618 \times 535$~mas (PA $=29\fdg8$) for $^{12}$CO and $636 \times 563$~mas (PA $=47\fdg0$) for $^{13}$CO. The rms noise in line-free channels was $3.5~\mathrm{mJy~beam^{-1}}$ for $^{12}$CO and $3.2~\mathrm{mJy~beam^{-1}}$ for $^{13}$CO.

\subsection{Archival Near-infrared Image}\label{sec:near_infra}

We use an archival near-infrared scattered-light image of the V1094~Sco disk obtained with VLT SPHERE in the $H$ band, originally presented by \citet{garufi_disks_2020}. We adopt the polarimetric $Q_{\Phi}$ image as the primary tracer of scattered-light from micron-sized grains. The observations were acquired with a coronagraph (mask diameter of $0\farcs185$), and we therefore treat the innermost region as unreliable for quantitative analysis.

For comparison with the ALMA continuum morphology, we perform two minimal processing steps. First, we compute the deprojected stellocentric radius in the disk plane using the disk geometry derived from the ALMA analysis (Section~\ref{sec:results_morphology_geometry}). Second, we construct a $r^{2}$-scaled version of the $Q_{\Phi}$ image. The resulting images and radial profiles, and their comparison with the millimeter continuum substructures, are presented in Section~\ref{sec:results_nir_mm}.

\section{Results}\label{sec:analysis_results}
In this section, we present the observational results that characterize the morphology, geometry, kinematics, and radial extent of the V1094 Sco disk. We first define the continuum-based disk geometry used throughout the work, then summarize the CO kinematic signatures, and finally quantify the disk radial extent and continuum substructures. A comparison with archival near-infrared scattered-light data is presented at the end of this section.

\subsection{Continuum Morphology and Disk Geometry}\label{sec:results_morphology_geometry}

\begin{figure*}[ht]
\centering
\includegraphics[width=0.98 \textwidth]{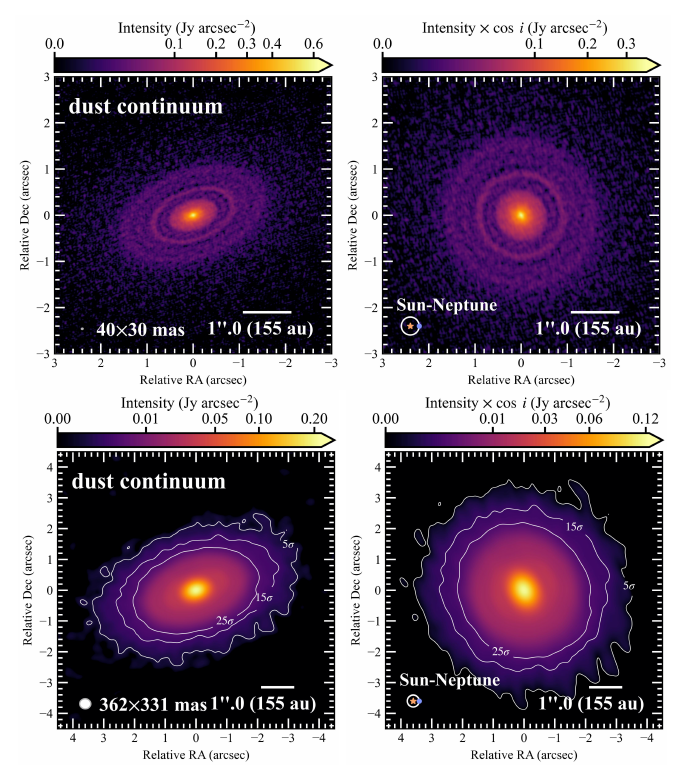}
\caption{Gallery of ALMA Band~6 ($1.3$~mm) dust continuum images of the V1094~Sco disk, reconstructed by PRIISM imaging. Top: Dust continuum distribution and its deprojected counterpart using the PRIISM model image. The image is originally defined in units of $\rm Jy~pixel^{-1}$ and is converted into $\rm Jy~arcsec^{-2}$ to allow a direct comparison with the PRIISM restored image presented in the bottom panels. The filled white ellipse denotes the effective spatial resolution $\theta_{\rm eff}$ estimated from an artificial point-source injection method. Bottom: Restored dust continuum image and its deprojected counterpart, produced by convolving the PRIISM model image with an elliptical Gaussian that represents the main lobe of the synthesized beam and subsequently adding the dirty residual map. White contours correspond to $[5,\,15,\,25]\times\sigma_{\rm noise}$, where $\sigma_{\rm noise}=23~\mu\rm Jy~beam^{-1}$ is the RMS noise level measured in an emission-free region. The restored image is originally expressed in $\rm Jy~beam^{-1}$ and is here converted to $\rm Jy~arcsec^{-2}$. The reference scale corresponding to the Sun's (orange) and Neptune's (blue) orbits ($r= 30$ au) is overplotted in the lower left corner of the deprojected images.}
\label{fig:priism_images}
\end{figure*}

Figure~\ref{fig:priism_images} shows the ALMA Band~6 dust continuum images reconstructed with $\tt PRIISM$. The top panels present the high-resolution PRIISM model image, while the bottom panels show the restored image used to visualize faint extended emission and to indicate the noise level of the image. The dust continuum images reveal a highly structured disk. Moving outward from the central stellar position, the emission exhibits an inner core ($r \leq 80$ au), followed by a sequence of narrow gaps and bright rings ($90 \leq r \leq 230$ au). The outer disk ($r \ge 300$ au) is characterized by an extended, low surface-brightness component at large radii. These features indicate that the disk is not only radially extended but also hosts fine-scale substructures superimposed on an otherwise smooth global profile.

To define a common reference frame for the analyses below, we infer the disk orientation by fitting an ellipse to the bright ring at $r=137$~au in the PRIISM model image, assuming the ring is intrinsically circular in the disk plane (following \citealt{yamaguchi_alma_2021}). We obtain a position angle $\mathrm{PA}=111\arcdeg.2\pm0\arcdeg.1$ and an inclination $i=54\arcdeg.7\pm0\arcdeg.1$. These values agree with 
$\sim5\%$ uncertainties \citep{van_terwisga_v1094_2018, vioque_alma_2025}.

\subsection{CO Emission and Rotation Signature}\label{sec:results_co_overview}

\begin{figure*}[ht!]
\centering
\includegraphics[width=0.98 \textwidth]{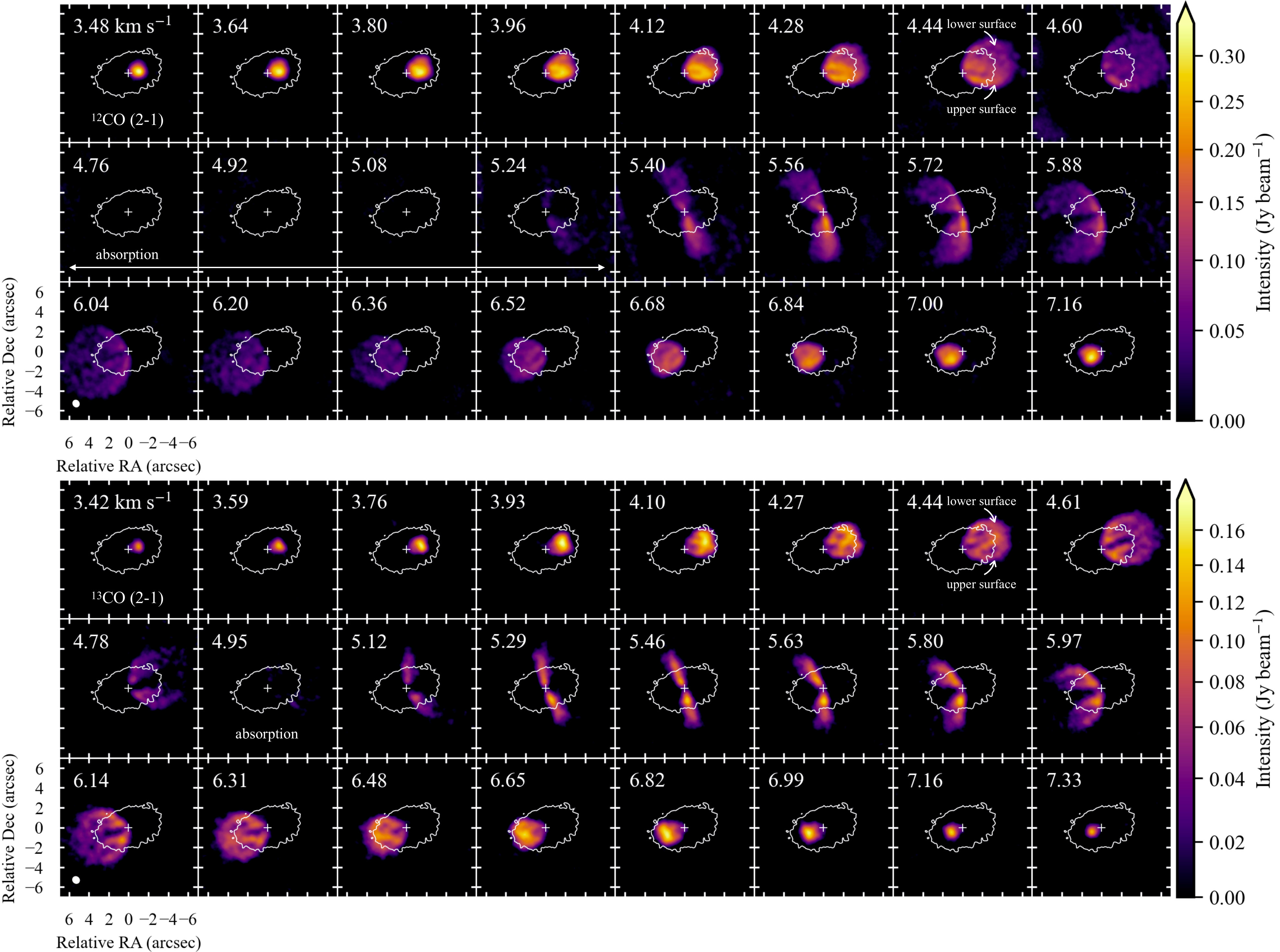}
\caption{Channel maps of the $^{12}$CO$(J=2-1)$ data cube (top panels) and the $^{13}$CO$(J=2-1)$ data cube (bottom panels). All images are reconstructed using CLEAN imaging. The corresponding line-of-sight velocity $v_{\rm LSR}$ in km s$^{-1}$ is indicated in white in each channel map. The $5\sigma$ contour of the PRIISM restored continuum image shown in Figure~\ref{fig:priism_images} is overlaid in white. The synthesized beam is shown in the lower left corner of each panel. Channels affected by absorption from foreground gas are marked with white annotations.}
\label{fig:channelmaps}
\end{figure*}

\begin{figure*}[ht!]
\centering
\includegraphics[width=0.98 \textwidth]{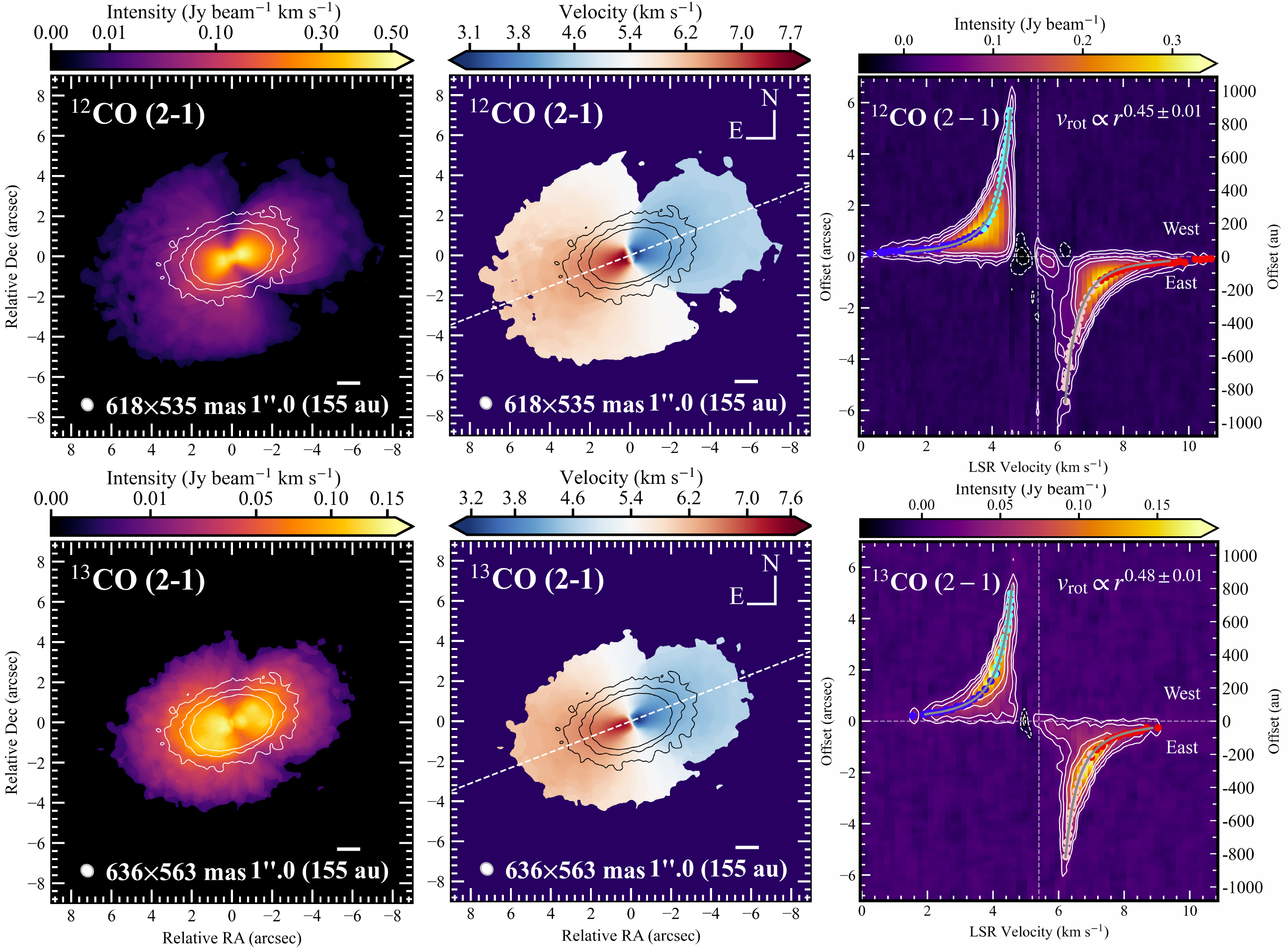}
\caption{Gallery of data cube maps of the $^{12}$CO$(J=2-1)$ emission (top panels) and the $^{13}$CO$(J=2-1)$ emission (bottom panels) in the V1094~Sco disk. All maps are imaged with CLEAN. Left: Velocity integrated intensity (moment 0) maps. Middle: Velocity field (moment 1) maps. The white dashed lines indicate the dust disk major axis with a position angle of $\mathrm{PA}=111\fdg2$. Contours of the PRIISM restored continuum image shown in Figure~\ref{fig:priism_images} are overlaid on the moment maps. Right: Position-velocity diagrams, extracted along the dust disk major axis indicated by the white dashed lines in the moment 1 maps. Contour levels are $[5, 10, 20, 30] \times \sigma$, where $\sigma$ denotes the RMS noise of each line image; dashed contours represent negative intensities at the same absolute levels. Bluish and reddish markers indicate representative blueshifted and redshifted data points, respectively, derived using \texttt{SLAM}. Light colored (cyan and pink) and thick colored (blue and red) markers correspond to measurements extracted from profiles along the velocity and positional axes, respectively. The gray solid curves show the best-fit single power-law models to the rotational profiles.}
\label{fig:momentmaps}
\end{figure*}

Figure~\ref{fig:channelmaps} presents channel maps of the CO isotopologue cubes, showing the characteristic butterfly pattern expected for a rotating disk. Emission from both the upper and lower disk surfaces is detected over $|v_{\rm LSR}-v_{\rm sys}|<1.6~\mathrm{km~s^{-1}}$, where $v_{\rm LSR}$ is the line-of-sight velocity and $v_{\rm sys}=5.4~\mathrm{km~s^{-1}}$ is the disk's systemic velocity (see Section~\ref{sec:rotation_curve}). The relative placement of the two emitting surfaces indicates that the near side of the disk is to the north and that the rotation is counterclockwise on the sky. Both isotopologues show absorption close to $v_{\rm sys}$, attributable to foreground Lupus~3 material at $v_{\rm LSR}\sim 4~\mathrm{km~s^{-1}}$ \citep{tachihara_13co_1996, Hara1999PASJ}.

The left and middle panels of Figure~\ref{fig:momentmaps} show the velocity integrated intensity (moment~0) and intensity weighted velocity (moment~1) maps constructed from the same cubes. The continuum contours overlaid on the moment maps show that the CO emission extends well beyond the millimeter dust, with $^{12}$CO tracing the broadest surface-brightness distribution and $^{13}$CO remaining more centrally concentrated. The moment~1 maps exhibit an ordered red to blue velocity gradient along the disk's major axis. To isolate disk emission in the presence of noise and foreground absorption, the moment maps are calculated by integrating emission detected above $3\sigma$ level and are derived from Keplerian-masked cubes using \texttt{Keplerian Mask Generator}\footnote[2]{$\texttt{Keplerian Mask Generator}$ is a publicly available tool developed by R.~Orihara to generate Keplerian masks for protoplanetary disk data cubes, with customizable disk and observation parameters: \url{https://github.com/rorihara/Keplerian_Mask_Generator}.}. Together with the four narrow continuum ring-gap pairs, these CO isotopologue moment maps show that V1094~Sco hosts strong substructures in the solids while retaining an exceptionally extended gaseous disk.

The right panel of Figure~\ref{fig:momentmaps} shows position-velocity (PV) diagrams extracted along the dust continuum major axis (PA $=111\fdg2$). The diagrams exhibit the characteristic signature of differential rotation in both $^{12}$CO and $^{13}$CO emission: in each tracer, the locus of peak emission shifts to larger $|v_{\rm LSR}-v_{\rm sys}|$ at smaller projected radii, as expected for a velocity field dominated by the gravitational potential of the central star \citep[e.g.,][]{aso_alma_2015, yen_signs_2017, sai_disk_2020}. No distinct kinematic component attributable to an infalling envelope is detected at the current sensitivity and spatial resolution. We also verify that moment maps constructed without Keplerian masking show the same ordered red-to-blue gradient across the disk and do not reveal any kinematic component with a steeper, non-Keplerian rotational profile. This is consistent with the absence of an envelope reported from ASTE observations, in which the $^{13}$CO $(J=3–2)$ column density toward V1094~Sco is well below the Lupus~3 cloud average \citep{tsukagoshi_detection_2011}. A quantitative analysis of the rotational velocity field is described in Section~\ref{sec:rotation_curve}.

\subsection{Disk radial extent from dust and gas tracers}\label{sec:results_disksize}

\begin{figure*}[t]
\centering
\begin{minipage}{0.95\textwidth}
   \centering
    \includegraphics[width=\linewidth]{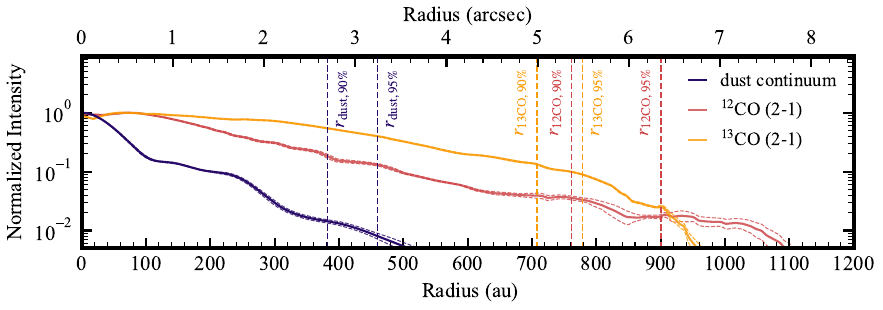}
    \caption{Comparison of the azimuthally averaged radial intensity profiles of the dust continuum (purple; extracted from the restored image), $^{12}$CO (red), and $^{13}$CO (orange). The intensity profiles are normalized to their respective radial peak intensities. The vertical dashed lines mark the disk radii enclosing $90\%$ and $95\%$ of the integrated flux. The radial profiles are interpolated onto a radial grid with 0.1~au spacing using {\tt interpolate.interp1d} from {\tt SciPy}. The uncertainty in the averaged intensity, $\hat{\sigma}_I$, is calculated as the standard error of the mean within each concentric ring, $\hat{\sigma}_I = \sigma_I / \sqrt{N_R}$, where $\sigma_I$ is the azimuthal brightness dispersion and $N_R = 2\pi r_i / \langle \theta \rangle$ is the number of independent resolution elements at radius $r_i$. Here, $\langle \theta \rangle$ denotes the geometric mean of the spatial resolution. The light shading indicates the error of the mean at each radius ($\hat{\sigma}_{I}$), although its amplitude is negligibly small across the entire radial range. For comparison, the azimuthal standard deviation ($\sigma_{I}$) is shown by the dashed curves.}
    \label{fig:disk_radius}
\end{minipage}
\begin{minipage}{0.95\textwidth}
    \centering
    \includegraphics[width=\linewidth]{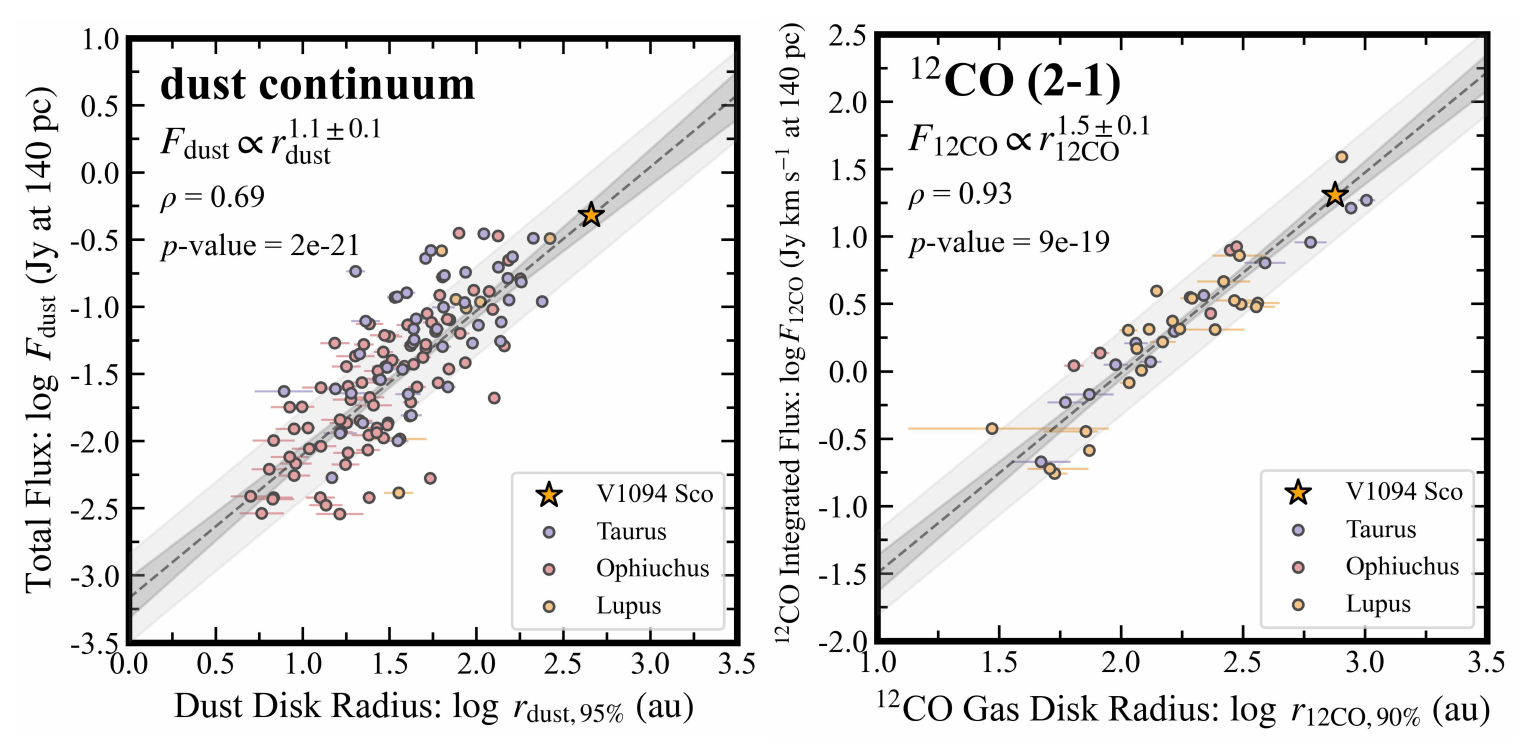}
\caption{Left: Relation between the Band~6 millimeter continuum flux density $F_{\rm dust}$, scaled to a distance of 140~pc (i.e., $F_{\rm dust}\times(d/140)^2$), and the dust disk radius $r_{\rm dust,95\%}$. The orange star symbol denotes V1094~Sco, while colored circles represent Class~II disks in Taurus (purple), Ophiuchus (red), and Lupus (orange). All disk radii are measured in a homogeneous manner across the samples and are defined as the radii enclosing $95\%$ of the total flux density, as derived using the curve-of-growth method. Right: Same relation as in the left panel, but using the $^{12}$CO (J=2-1) line emission and the gas disk radii enclosing $90\%$ of the integrated $^{12}$CO flux. In each panel, the black dashed line indicates the median scaling relation obtained from Bayesian linear regression, and the dark gray shaded region represents the $68\%$ confidence interval around the median relation. The light gray shaded region corresponds to the $\pm 1\sigma$ dispersion of the data points around the best-fit relation, obtained by fitting a Gaussian to the histogram of residuals (data minus model). The best-fit linear regression parameters, the Pearson correlation coefficient ($\rho$), and the associated $p-$value are listed in the upper left corner of each panel.}
\label{fig:diskradius_flux}
\end{minipage}
\end{figure*}

\begin{table*}[htbp]
\caption{Disk radii and total fluxes \label{tab:disk_radii}}
\begin{tabular*}{\linewidth}{@{\extracolsep{\fill}}lcccccccc@{}}
\toprule
Tracer
& $r_{68\%}$ 
& $r_{68\%}$ 
& $r_{90\%}$ 
& $r_{90\%}$ 
& $r_{95\%}$ 
& $r_{95\%}$ 
& $\sigma_r$
& Flux \\
& (arcsec)
& (au)
& (arcsec)
& (au)
& (arcsec)
& (au)
& (arcsec, au)
&  \\
\midrule
Dust continuum 
& 1.52 & 236 
& 2.47 & 382 
& 2.97 & 457 
& 0.15, 23 
& $394$~mJy \\
$^{12}$CO ($J=2-1$)
& 2.87 & 444 
& 4.92 & 762 
& 5.82 & 900 
& 0.24, 38 
& $16562$~mJy~km~s$^{-1}$ \\
$^{13}$CO ($J=2-1$)
& 3.24 & 501 
& 4.57 & 707 
& 5.03 & 778 
& 0.25, 39 
& $10823$~mJy~km~s$^{-1}$ \\
\bottomrule
\end{tabular*}
\tablecomments{
Disk radii enclosing different fractions of the total flux for the dust continuum and CO isotopologue emission in V1094~Sco. The radii are measured using the curve-of-growth method. The quoted uncertainty $\sigma_r$ represents the radial uncertainty arising from the spatial resolution 
$\sigma_{r} = \langle\theta\rangle /(2\sqrt{2\ln 2})$.}
\end{table*}

We quantify the disk radial extent using a curve-of-growth analysis applied consistently to the deprojected continuum image and to the CO isotopologue moment 0 maps. After deprojection to a face-on orientation using the continuum-based inclination and position angle, the cumulative flux enclosed within radius $r$ is
\begin{align}
F_{\nu}(r)
&
= 2\pi \int_{0}^{r} I_{\nu}(r^{\prime}) r^{\prime} \, \mathrm{d}r^{\prime} \nonumber\\
&\simeq 2\pi \sum_{j=1} I_{\nu}(r_{j}) r_{j} \, \Delta r ,
\end{align}
where $I_{\nu}(r_{j})$ is the azimuthally averaged surface brightness in each radial bin. Operationally, this corresponds to measuring the enclosed flux within successively larger circular apertures until convergence to the total flux. We define characteristic radii as the radii enclosing fixed fractions ($68\%$, $90\%$, and $95\%$) of the total emission and compare dust and gas radii in a homogeneous manner.

Throughout this analysis, we define the azimuthal selection as a wedge centered on the disk major axis with full opening angle $\Delta\phi$. With this convention, $\Delta\phi=180\arcdeg$ corresponds to azimuthal averaging over the full deprojected disk, while smaller values progressively exclude the regions near the minor axis.

Millimeter continuum emission traces pebble-sized grains that have settled toward the disk midplane \citep[e.g.,][]{chung_sma_2024}. We therefore apply the above procedure to the restored continuum image using the full azimuthal range ($\Delta\phi=180\arcdeg$), assuming that the continuum emission originates from a geometrically thin layer.

For the gas disk, we apply the same analysis to the CO isotopologue moment~0 maps but restrict the measurement to the redshifted eastern side of the disk, where foreground absorption is negligible. In addition, only emission within a narrow wedge of $\Delta\phi=30\arcdeg$ is retained. This choice mitigates systematic inflation of the inferred gas radii caused by enhanced vertical projection effects near the minor axis when performing wider azimuthal averaging (see Appendix~\ref{apx:gasdisksize_bias} for more details). The resulting dust and gas radii are summarized in Table~\ref{tab:disk_radii}.

Figure~\ref{fig:disk_radius} compares the azimuthally averaged radial intensity profiles of the dust continuum and CO isotopologue emission, extracted using the azimuthal selections described above. The radii enclosing $90\%$ of the integrated flux are $r_{\rm d,90\%}=382$~au for the continuum, $r_{{}^{13}{\rm CO},90\%}=707$~au for $^{13}$CO, and $r_{{}^{12}{\rm CO},90\%}=762$~au for $^{12}$CO. The ordering $r_{{}^{12}{\rm CO},90\%} > r_{{}^{13}{\rm CO},90\%} \gg r_{\rm d,90\%}$ reflects the distinct disk layers and optical-depth regimes traced by each component. Owing to its high optical depth, $^{12}$CO remains detectable at low column densities and predominantly traces warm molecular layers at elevated heights. In contrast, $^{13}$CO probes deeper regions with lower optical depth and becomes sensitivity limited at smaller radii \citep[e.g.,][]{nomura_high_2021}. The more compact continuum emission indicates that millimeter-sized grains are relatively concentrated, consistent with grain growth and radial drift, which deplete millimeter-sized solids in the outer disk while the molecular gas remains extended \citep[e.g.,][]{trapman_gas_2019}.

Figure~\ref{fig:diskradius_flux} compares disk size and integrated flux for the dust continuum and $^{12}$CO emission, measured consistently for nearby Class~II samples and scaled to a common distance of 140~pc.\footnote[3]{Using Bayesian linear regression with \texttt{Linmix} \citep{kelly_aspects_2007} on logarithmic scales, we find $\log F_{\rm dust} \propto (1.1 \pm 0.1)\,\log r_{\rm dust}$ with a residual dispersion of $0.33 \pm 0.01$~dex after scaling fluxes to 140~pc. For $^{12}$CO, we find $\log F_{{}^{12}{\rm CO}} \propto (1.5 \pm 0.1)\,\log r_{{}^{12}{\rm CO},90\%}$ with a residual dispersion of $0.31 \pm 0.05$~dex.} In the continuum, V1094~Sco lies at the upper end of the size distribution. Its dust radius, defined at the $95\%$ flux level, exceeds those of all disks in Taurus \citep{yamaguchi_alma_2024}, Ophiuchus \citep{shoshi2025a}, and Lupus \citep{vioque_alma_2025, Huang2018ApJ}, and this result persists for alternative radius definitions (i.e., dust disk radii defined at the $68\%$ and $90\%$ flux levels) and additional nearby samples \citep[e.g.,][]{tazzari_multi-wavelength_2020, hendler_evolution_2020}. The gas disk is also extreme: the $^{12}$CO radius is among the largest reported for Class~II systems in these regions \citep{long_gas_2022, trapman_alma_2025}, placing V1094~Sco at the upper boundary of the observed distribution. Despite its extreme size and brightness, V1094~Sco follows the same empirical size versus flux trends as the broader Class~II population in both the dust continuum and $^{12}$CO, occupying the high radius and high flux end of these relations rather than deviating from them.

\subsection{Continuum substructures}\label{sec:results_substructures}

\begin{table*}[htbp]
\caption{Properties of disk substructures\label{tab:disk_substrucutures}}
\begin{tabularx}{\linewidth}{ccccccccc}
\toprule
Type & Label & Gap & Ring & Inflection & Gap Width & Norm Gap Width & Gap Depth \\
 & & $r_{\rm gap}$ (au,mas) & $r_{\rm ring}$ (au,mas) & $r_{\rm inf}$ (au,mas) & $\Delta_{\rm I, unit}$ (au,mas) & $\Delta_{\rm I}$ & $\delta_{\rm I}$\\
 (1) & (2) & (3) & (4) & (5) & (6) & (7) & (8) \\
\midrule
Shoulder & I20       & $\cdots$     & $\cdots$     & 20.2 (130) &$\cdots$ & $\cdots$ & $\cdots$\\
Shoulder & I38       & $\cdots$     & $\cdots$     & 37.6 (243) &$\cdots$ & $\cdots$ & $\cdots$ \\
Shoulder & I52       & $\cdots$     & $\cdots$     & 51.7 (334) &$\cdots$ & $\cdots$ & $\cdots$\\
Shoulder & I65       & $\cdots$     & $\cdots$     & 64.8 (419) &$\cdots$ & $\cdots$ & $\cdots$\\
Ring-gap & D95/B109  & 95.2 (616)   &  108.8 (703) & $\cdots$   & 10.6 (68.6) & 0.11 & $1.69\pm 0.03$ \\
Ring-gap & D119/B137 & 119.4 (772)  & 137.4 (889)  & $\cdots$   & 17.8 (114.8) & 0.14 & $3.66\pm 0.06$ \\
Disk-skirt & I159      & $\cdots$     & $\cdots$     & 159.4 (1030) &$\cdots$ & $\cdots$ & $\cdots$\\
Ring-gap & D171/B187 & 171.2 (1106) & 187.0 (1209) & $\cdots$   & 17.3 (112.0) & 0.10 & $1.40\pm0.02$ \\
Ring-gap & D205/B231 & 205.1 (1326) & 231.1 (1494) & $\cdots$   & 18.5 (119.7) & 0.09 & $1.64\pm0.02$ \\
Disk-skirt & I318    & $\cdots$     & $\cdots$     & 318.4 (2058) &$\cdots$ & $\cdots$ & $\cdots$\\
\bottomrule
\end{tabularx}
\tablecomments{Column descriptions:
(1) Disk substructure type; (2) Substructure label; (3) Radial gap location in astronomical unit (au) and millimeter-arcsecond (mas); (4) Radial ring location in au and mas. (5) Radial inflection location in au and mas; (6) Gap width in au; (7) Normalized gap width; (8) Gap depth. The uncertainties of the gap properties are $ 1\sigma$ and do not account for the uncertainty in the distance to the source.}
\end{table*}

\begin{figure}[ht!]
\centering
\includegraphics[width=0.44 \textwidth]{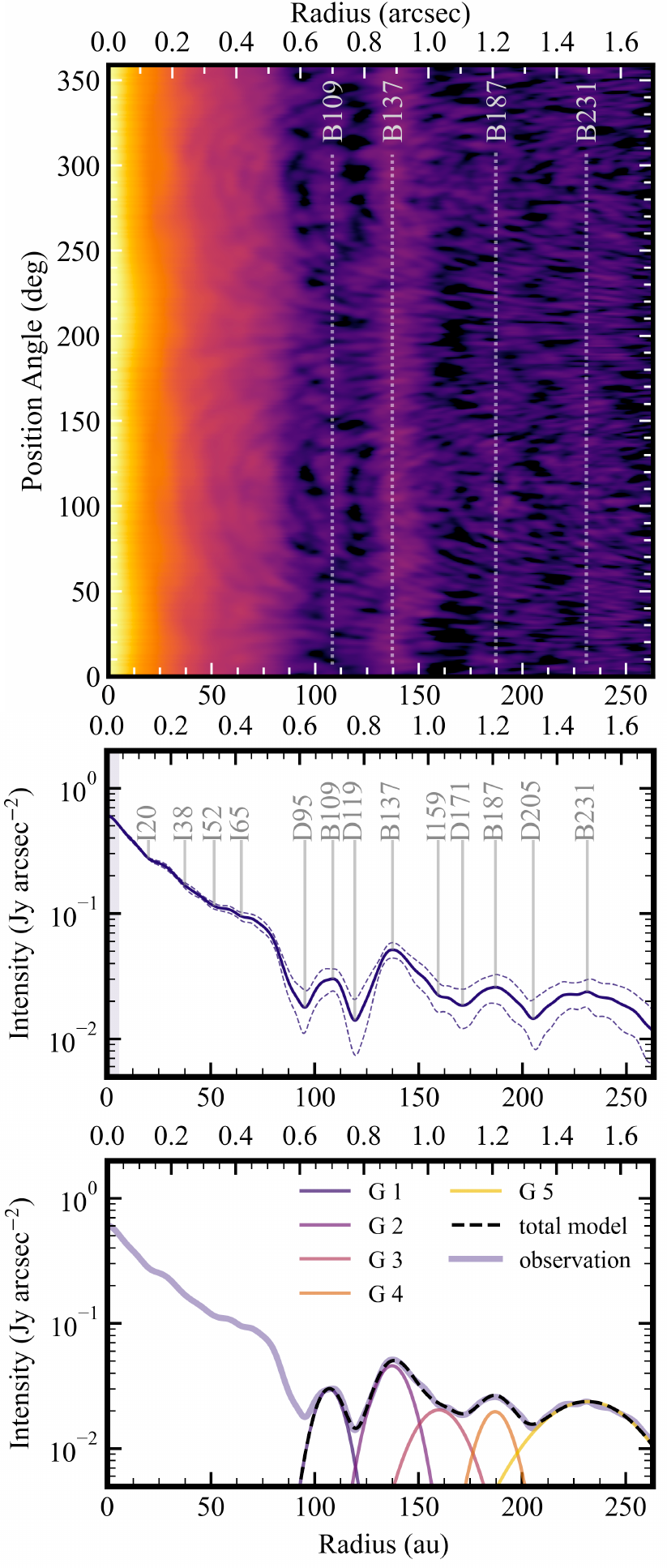}
\caption{Deprojected PA profile and azimuthally averaged radial intensity profiles, derived from the PRIISM model image. The profile configuration is the same as in Figure \ref{fig:disk_radius}. Vertical gray lines mark the locations of gaps ($D$), rings ($B$), and inflection points ($I$). The panels show the PA profile (top) and the intensity profiles (middle) with the best-fit multi--Gaussian model overplotted (bottom). The vertical purple band on the left in the middle panel shows the corresponding effective resolution scale.}
\label{fig:priism_radialprofile}
\end{figure}

\begin{figure}[htp!]
\centering
\includegraphics[width=0.45 \textwidth]{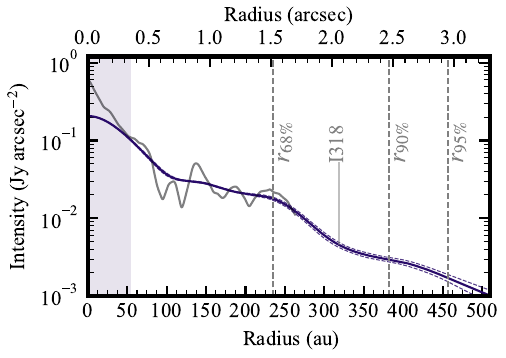}
\caption{Deprojected and azimuthally averaged radial intensity profile (purple curve) on a logarithmic scale, derived from the PRIISM restored image. The profile follows the same framework as in Figure~\ref{fig:disk_radius}. Vertical dashed lines mark the dust disk radii determined by the curve-of-growth method, while vertical gray lines indicate the locations of the inflection points ($I$) regarded as disk-skirt feature. For comparison, the ring-gap area ($r_{\rm d} \leq 1\farcs7$) on the intensity profile obtained from the PRIISM model image is overplotted as a gray curve.}
\label{fig:priism_radialprofile_restored}
\end{figure}

Figures~\ref{fig:priism_radialprofile} and \ref{fig:priism_radialprofile_restored} present azimuthally averaged radial intensity profiles, $I_{\nu}(r)$, extracted from the PRIISM model and restored images after deprojection to a face-on orientation. The profile $I_{\nu}(r)$ is computed by averaging the deprojected emission within a $60\arcdeg$ wedge centered on the disk's semi major axis, excluding azimuths near the minor axis where projection effects can bias both the apparent contrast and the inferred radial locations of substructures.

The radial intensity profile derived from the model image in Figure~\ref{fig:priism_radialprofile} declines smoothly, with only weak modulations inside $\sim70$~au. At larger radii, the profile becomes more structured, showing a sequence of narrow peaks and troughs superposed on an overall decline. This behavior indicates multiple localized dust concentrations embedded within a global radial gradient. The radial intensity profile derived from the restored image (purple curve in Figure~\ref{fig:priism_radialprofile_restored}) exhibits a change in slope at $r_{\rm d}\sim300$~au, marking a transition from the ring-gap complex to a faint extended emission dominating the outer disk.

To identify and classify substructures, we adopt the derivative-based scheme described in \citet{yamaguchi_alma_2024}. In this approach, local extrema in the slope and curvature of $I_{\nu}(r)$ are used to identify rings (``B'' for bright), gaps (``D'' for dark), and inflection points (``I''). The features are labeled with a number indicating their location in astronomical units. The mathematical definitions and implementation details are provided in Appendix~\ref{apx:substructure_definition}. The measured locations, widths, and depths of the gap features are summarized in Table~\ref{tab:disk_substrucutures}. We confirm that all detected gaps are spatially resolved, as the widths of all gaps are larger than the geometric mean of the PRIISM effective resolution (i.e., $\langle\theta_{\rm eff}\rangle\simeq 5~ \rm au$). Below, we summarize the resulting phenomenology.

We identify three categories of continuum substructures, arranged in a clear radial hierarchy. At small radii ($r_{\rm d}<70$~au), the profile contains four shallow curvature features (I20, I38, I52, and I65). These ``shoulders'' appear as gentle modulations on the monotonic inner-disk decline and do not form complete ring-gap pairs.

At intermediate radii, the disk exhibits a distinctive W-shaped morphology consisting of two closely spaced ring-gap pairs (D95/B109 and D119/B137). The outer gap D119 is deeper and wider than the inner gap D95 by roughly a factor of two.

Further out, between $171$ and $230$~au in radius, we detect two additional ring-gap pairs (D171/B187 and D205/B231). In contrast to the intermediate W-shaped structure, the two outer gaps have comparable depths and widths within $\sim10\%$. The outermost pair (D205/B231) lies at an exceptionally large stellocentric distance compared to ring-gap structures reported in current Class~II samples \citep{Huang2018ApJ, cieza_ophiuchus_2021, yamaguchi_alma_2024, huang_high-resolution_2024, guerra-alvarado_high-resolution_2025}.

Beyond $r_{\rm d}\gtrsim230$~au, the radial profile only seen in the restored image transitions into a broad outer component, characterized by an inflection point (I318) followed by a low-contrast exterior ridge. We also identify a similar inflection feature (I159) exterior to the prominent ring (B137) on the model image. We refer to those features as a ``disk skirt'' which are regarded as a gradual transition from the ring to the diffuse outer dust reservoir.

The four rings are moderately resolved in the radial direction relative to the effective spatial resolution. To quantify characteristic ring widths in a uniform manner, we fit the radial intensity profile over $90<r<265$~au with a simple model composed of $N=5$ Gaussian components,
\begin{align}
I^{\rm gauss}_{\nu}(r)=\sum_{i=1}^{N} A_{i}\,
\exp \left[-\frac{1}{2}\left(\frac{r-\mu_{i}}{w_{d,i}}\right)^{2}\right],
\end{align}
\noindent
where $A_i$, $\mu_i$, and $w_{d,i}$ are the amplitude, peak radius, and width of each component. This parameterization is used solely to measure ring widths and their uncertainties; reproducing the full profile with multiple components is a byproduct of the fit. We note that one component (G4 in the bottom panel of Figure~\ref{fig:priism_radialprofile}) represents the broad, low-contrast emission associated with the disk-skirt transition. We infer the parameters with a Markov Chain Monte Carlo (MCMC) analysis\footnote[4]{We adopted uniform priors within physically reasonable bounds and sampled the posterior distribution using 100 walkers evolved for 3000 steps, discarding the first 1000 as burn in. The likelihood function assumes independent Gaussian errors and evaluates the agreement between the observed and modeled intensities using the measured uncertainties at each radial position. The final parameter estimates are given by the median of the posterior samples, with the corresponding uncertainties being the 16th and 84th percentiles.} using the \texttt{emcee} ensemble sampler \citep{Foreman2013}. The best-fit model is shown in the bottom panel of Figure~\ref{fig:priism_radialprofile}.

To estimate intrinsic ring widths, we deconvolve the fitted Gaussian widths assuming Gaussian beam smearing. The deconvolved width is $\hat{w}_{d}=\sqrt{w_{d}^{2}-\sigma_{b}^{2}}$, where $\sigma_b$ is the geometric mean of the beam standard deviations for the effective resolution. Deconvolved widths and best-fit parameters are summarized in Table~\ref{tab:prop_disk_substructures}. Except for the outermost component, the inferred $\hat{w}_{d}$ values are typically $\sim10$~au. These widths are used to assess dust-trapping efficiency (Section~\ref{sec:dust_trapping}) and to constrain turbulent viscosity (Section~\ref{sec:alpha_constraint}).

\begin{table}[ht!]
\centering
\caption{best-fit ring properties}
\label{tab:prop_disk_substructures}
\setlength{\tabcolsep}{3.8pt} 
\begin{tabular}{ccccc}
\hline
Label & Amp $A$ & Radius $\mu$ & Width ${w}_{d}$ & Deconv $\hat{w}_{d}$\\
 & ($\rm mJy~asec^{-2}$) & (au) & (au) & (au)\\
\hline
G1 & $29.9^{+0.1}_{-0.1}$ & $107.02^{+0.05}_{-0.02}$ & $7.48^{+0.01}_{-0.03}$ & $7.13^{+0.01}_{-0.03}$\\
G2 & $45.8^{+1.7}_{-1.7}$ & $137.30^{+0.29}_{-0.20}$ & $9.14^{+0.22}_{-0.22}$ & $8.85^{+0.23}_{-0.23}$\\
G3 & $20.3^{+0.7}_{-0.7}$ & $159.92^{+1.18}_{-1.00}$ & $12.98^{+1.20}_{-1.28}$ & $12.77^{+1.22}_{-1.30}$\\
G4 & $19.7^{+1.2}_{-1.0}$ & $187.00^{+0.60}_{-0.62}$ & $8.71^{+0.21}_{-0.35}$ & $8.40^{+0.21}_{-0.37}$\\
G5 & $0.024^{+0.001}_{-0.001}$ & $231.46^{+0.63}_{-0.59}$ & $24.05^{+0.58}_{-0.54}$ & $23.94^{+0.58}_{-0.55}$\\
\hline
\end{tabular}
\end{table}

\subsection{Near-infrared scattered-light Comparison}\label{sec:results_nir_mm}

\begin{figure*}[ht]
\centering
\includegraphics[width=0.98 \textwidth]{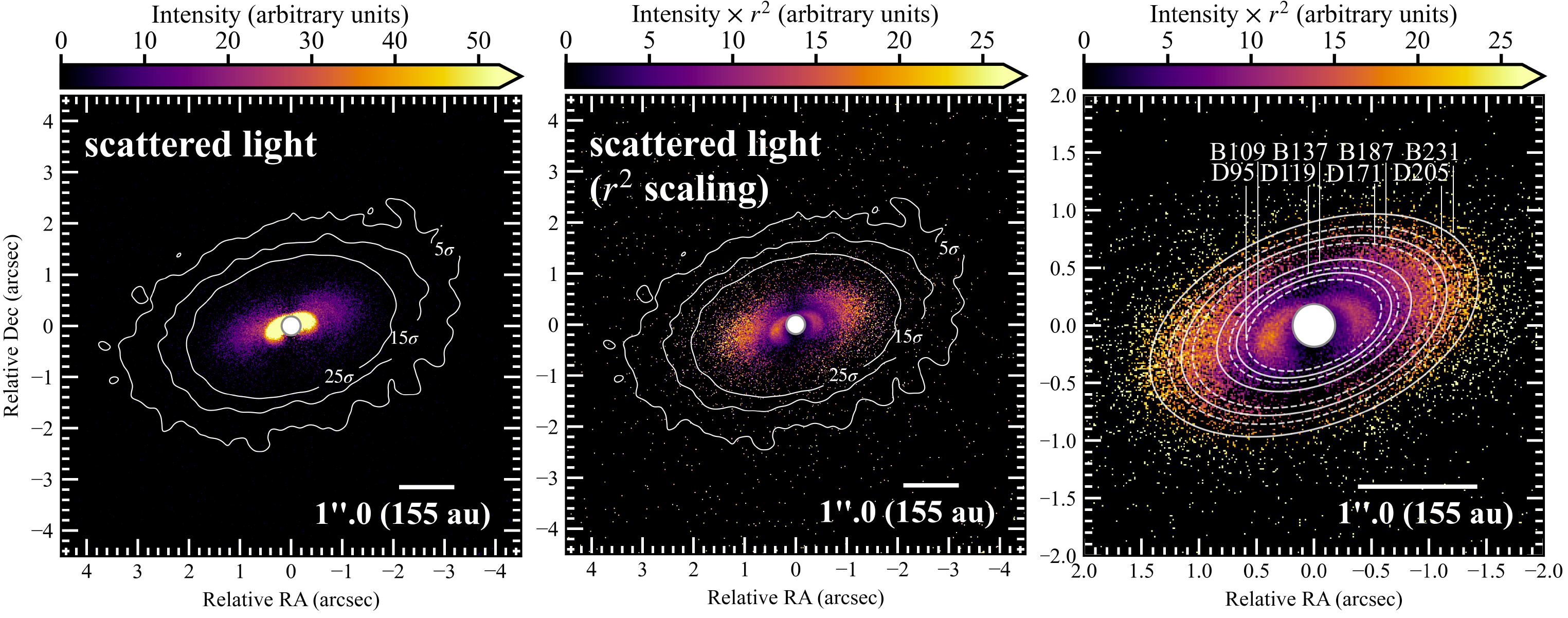}
\caption{Near-infrared scattered-light images of the V1094~Sco disk obtained with VLT/SPHERE in the $H$ band ($\lambda = 1.6~\mu$m; \citealt{garufi_disks_2020}). 
Left: Polarimetric $Q_{\Phi}$ image tracing scattered-light from micron-sized dust grains. White contours show the ALMA Band~6 dust continuum emission ($\lambda = 1.3$~mm; this work), which traces millimeter-sized grains; the contour levels are identical to those in Figure~\ref{fig:priism_images}. 
Middle: Same as the left panel, but with the scattered-light image scaled by $r^{2}$ to compensate for the radial dilution of stellar irradiation. 
Right: Close-up view of the $r^{2}$-scaled image, overlaid with the locations of dust continuum rings (solid lines) and gaps (dashed lines). The filled white circle indicates the SPHERE coronagraph with a diameter of $0\farcs185$.}
\label{fig:scattered_image}
\end{figure*}

We reanalyze an archival near-infrared scattered-light image of the V1094~Sco disk presented in \citet{garufi_disks_2020}. The left panel of Figure~\ref{fig:scattered_image} shows the original polarimetric $Q_{\Phi}$ image obtained with VLT/SPHERE in the $H$ band. The scattered-light emission is dominated by a bright inner region extending to $\sim 80$~au, followed by a sharp drop in intensity and a more gradually declining outer region. A localized dip along the minor axis is also identified, which may be attributed to residual stellar polarization rather than to an intrinsic disk feature.

\begin{figure}[ht]
\centering
\includegraphics[width=0.48 \textwidth]{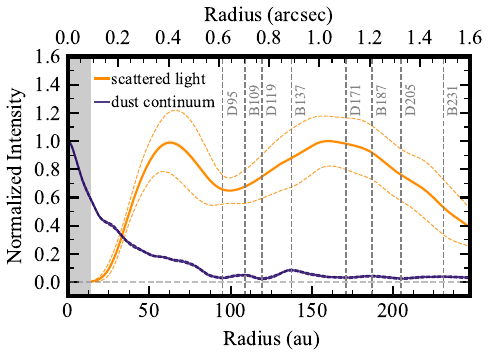}
\caption{Comparison of azimuthally averaged radial intensity profiles of the $r^{2}$-scaled scattered-light image (orange) and the dust continuum image (purple; same as Figure~\ref{fig:priism_radialprofile}). The radial profile extraction follows the same procedure as in Figure~\ref{fig:disk_radius}, and each profile is normalized to the peak intensity. The scattered-light image was convolved with a Gaussian kernel of FWHM $0\farcs1$ to suppress pixel-to-pixel fluctuations and produce a smoother radial profile. Azimuthal averaging was performed within a $60\arcdeg$ wedge centered on the disk semi major axis. The vertical dashed lines indicate the radial locations of disk gaps. The gray area indicates the area of the coronagraph. The profile at radii close to the coronagraph mask should be interpreted with caution because residual instrumental and stellar-polarization systematics can affect the $r^{2}$-scaled image.}
\label{fig:scattered_image_radialprofile}
\end{figure}

The appearance changes after applying an inclination-corrected $r^{2}$ scaling. The middle panel of Figure~\ref{fig:scattered_image} shows the resulting image and reveals a gap that is not evident in the original data. As shown in Figure~\ref{fig:scattered_image_radialprofile}, the radial location of this scattered-light gap coincides with the W-shaped substructure (D95--B109--D119) in the dust continuum. This is a relatively rare case in which substructures are detected at consistent radii in both near-infrared scattered-light and dust continuum emission \citep[e.g., TW Hya and HD 169142;][]{boekel_three_2017, bertrang_hd_2018}. Given that these tracers probe distinct dust populations and different vertical layers of the disk, their spatial coincidence suggests a common underlying physical origin. 

The correspondence is, however, not universal. Figure~\ref{fig:scattered_image_radialprofile} also shows that no scattered-light counterpart is detected at the location of the outer continuum gaps (D171 and D205). This wavelength-dependent behavior suggests that some substructures primarily affect the midplane distribution of millimeter-sized grains while leaving the disk surface traced by micron-sized particles largely unchanged. Such differences provide a diagnostic for distinguishing among competing mechanisms for ring and gap formation, discussed in Section~\ref{sec:ring-gap_origin}.

A further difference between the two tracers appears in their overall radial extents. Figure~\ref{fig:scattered_image} also shows that the millimeter continuum emission extends to larger radii than the near-infrared scattered-light. Such a mismatch is consistent with a vertically settled outer disk in which the scattering-surface becomes faint beyond the illumination front, for example, due to partial self-shadowing. Dust settling is expected when vertical stirring is inefficient, and settling induced self shadowing strongly suppresses scattered-light from the shadowed regions \citep{dullemond_effect_2004, garufi_sphere_2022}. This trend suggests the weak turbulence on the disk, which is discussed in Section~\ref{sec:alpha_constraint}.

We note that the innermost region ($r<0\farcs4$) is affected by the SPHERE coronagraph and by residual instrumental and stellar-polarization systematics, and the $r^{2}$ scaling can amplify such effects. We therefore interpret the radial profile interior to the coronagraph mask (and its immediate surroundings) with caution, and focus on substructures at larger radii where the disk signal is robust.

\section{Analysis}\label{sec:analysis}

In this section, we derive key physical properties of the V1094~Sco disk. We first refine the dynamical stellar mass from the CO kinematics. We then use the scattering-surface geometry to constrain the disk flaring and infer the disk's temperature. Finally, we derive dust-trapping diagnostics from the ring widths by comparing the intrinsic ring confinement scale to the local gas pressure scale height.

\subsection{Disk Kinematics and Dynamical Stellar Mass}\label{sec:rotation_curve}

With the PV diagrams shown in Figure~\ref{fig:momentmaps}, we quantify the rotational velocity fields by fitting a power-law model to the emission ridge using the \texttt{pvanalysis} module within the Spectral Line Analysis Modeling framework (\texttt{SLAM}; \citealt{aso_spectral_2024}). Ridge points are identified by taking one-dimensional cuts through the PV diagram and computing intensity weighted mean positions or velocities, following the procedure described in \citet{aso_spectral_2024}. Only emission above a $5\sigma$ threshold is used to define the ridge. When the emission is sufficiently spatially resolved, ridge fitting provides a robust estimate of the dynamical stellar mass \citep{aso_alma_2015}. We model the projected rotation speed as
\begin{equation}
\label{eq:rotation_fit}
v_{\mathrm{rot}} \equiv \left|v_{\mathrm{LSR}}-v_{\mathrm{sys}}\right|
= v_0\left(\frac{r}{r_0}\right)^{-p},
\end{equation}
where $p$ is the logarithmic slope, and $(v_0,r_0)$ set the normalization. In our fitting, $v_0$ is fixed to the mean velocity of the ridge points, while $(r_0,p,v_{\rm sys})$ are explored with the MCMC sampler using \texttt{emcee} in \texttt{SLAM}. 

The best-fit parameters are summarized in Table~\ref{tab:rotation_fit}. Both tracers yield a consistent systemic velocity of $v_{\rm sys}=5.40\pm0.01~\mathrm{km~s^{-1}}$. The inferred slopes are close to the Keplerian expectation ($p=0.5$), with $p=0.45\pm0.01$ for $^{12}$CO and $p=0.48\pm0.01$ for $^{13}$CO.

The dynamical stellar mass is then derived from the normalization of the rotation curve assuming Keplerian rotation ($p=0.5$), using
\begin{equation}
M_\star = \frac{r\,v_{\rm rot}^2}{G\,\sin^2 i}
\end{equation}
where $G$ is the gravitational constant and $i~(= 54\arcdeg.7)$ is the disk inclination. The stellar masses derived from the two CO isotopologues are summarized in Table~\ref{tab:rotation_fit}. The value inferred from $^{12}$CO is about $10\%$ lower than that inferred from $^{13}$CO. A natural explanation is that the optically thick $^{12}$CO emission preferentially traces higher molecular layers where the rotation becomes mildly sub-Keplerian owing to pressure support and the reduced stellar gravitational acceleration projected onto the disk midplane at finite height \citep{pinte_direct_2018}. We therefore adopt the $^{13}$CO based estimate, $M_{\star}=0.88~M_{\odot}$, as the representative dynamical stellar mass used throughout this work.

The formal MCMC uncertainties reported by \texttt{SLAM} (i.e., the central $68\%$ posterior interval; \citealt{aso_spectral_2024}) primarily quantify the statistical precision within the adopted likelihood function and parametric model. However, additional systematics can arise from inaccuracies in the assumed disk geometry (e.g., the emission height, position angle, and inclination), absorption by foreground cloud gas, and observational effects such as beam convolution and incomplete $uv$ sampling. To account for these effects in the stellar mass estimate used throughout this paper, we adopt a conservative $10\%$ systematic uncertainty, motivated by synthetic observations for numerical disk models with $M_{\star}>0.3~M_{\odot}$ and $i\sim50\arcdeg$ \citep{aso_testing_2020}.

\begin{table}[htbp]
\begin{center}
\caption{Results of the rotation curve fitting\label{tab:rotation_fit}}
\begin{tabular}{l@{\hskip 1mm}c@{\hskip 1mm}c@{\hskip 1mm}c@{\hskip 1mm}c@{\hskip 1mm}c}
\toprule
Line & \makebox[5mm][c]{$v_0$} & \makebox[7mm][c]{$r_0$} & \makebox[13mm][c]{$p$} & 
\makebox[13mm][c]{$v_{\rm sys}$} & \makebox[13mm][c]{$M_\ast$} \\
     & (km s$^{-1}$) & (au) & & (km s$^{-1}$) & ($M_\odot$) \\
(1) & (2) & (3) & (4) & (5) & (6) \\
\midrule
$^{12}$CO & 2.17 & $105 \pm 1$ & $0.45 \pm 0.01$ & $5.40 \pm 0.01$ & $0.83 \pm 0.08$ \\ 
$^{13}$CO & 1.73 & $174 \pm 2$ & $0.48 \pm 0.01$ & $5.40 \pm 0.01$ & $0.88 \pm 0.09$\\
\bottomrule
\end{tabular}
\end{center}
\tablecomments{
Column descriptions: (1) Molecular line used in the fit. (2)–(5) Parameters defined in Equation~\ref{eq:rotation_fit}. Here, $v_0$ denotes the average velocity of the data points, while the remaining three are free parameters in the fitting procedure. (6) Stellar dynamical mass inferred from $v_0$ under the assumption of Keplerian rotation (with the power-law index fixed at $p = 0.5$) and adopting the disk inclination of $54\arcdeg.7$.}
\end{table}

\subsection{Surface Geometry from Scattered Light}\label{sec:shape_disksurface}

\begin{figure*}[ht!]
\centering
\includegraphics[width=0.98\textwidth]{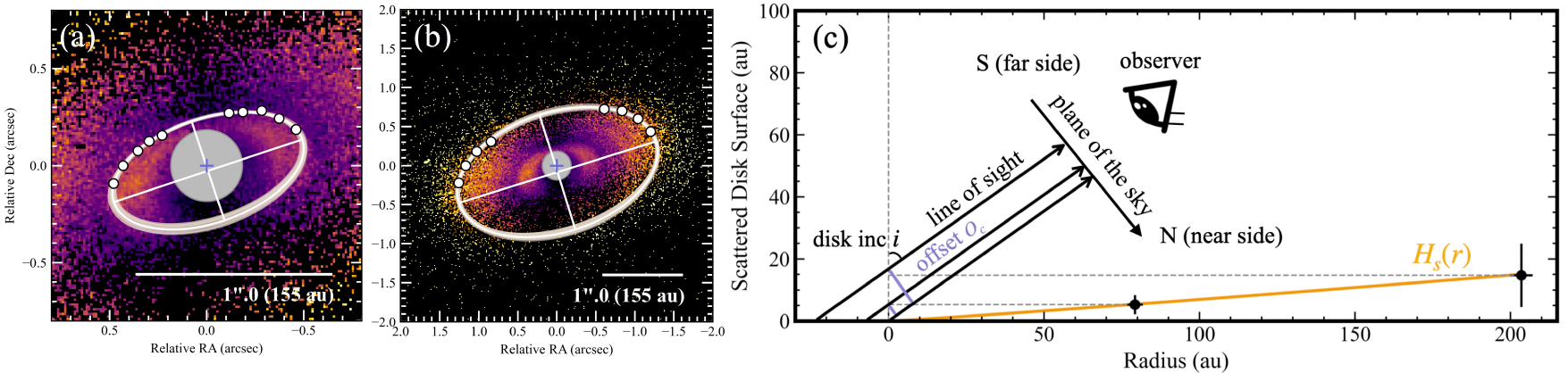}
\caption{Geometric determination of the scattering-surface height in the V1094~Sco disk.
The inclination-corrected $r^{2}$-scaled scattered-light images are shown with the best-fit ellipses overlaid (thick white curves), tracing the $\tau\sim1$ scattering-surface at a given radius for the (a) inner edge and (b) outer edge regions. The light white curves illustrate the distribution of ellipse solutions within the estimated $1\sigma$ uncertainties. White circles mark the radially averaged edge positions extracted from the scattered-light image at azimuthal intervals of $10\arcdeg$; only near-side (forward-scattering) points are used in the fits, while the far-side edge is excluded because of reduced back-scattered intensity. The blue cross indicates the stellar position. (c) Schematic illustration of the geometric reconstruction of the scattering-surface height $H_{\mathrm{s}}(r_{\rm sca,edge})$. Owing to disk inclination, the projected scattering-surface appears as an ellipse whose center is offset from the stellar position along the minor axis by an amount $O_{\rm c}$. This offset provides a direct measure of the scattering-surface height via simple geometry. The two measured heights at different radii, derived from panels (a) and (b), are used to constrain the radial shape of the disk surface shown in panel (c).}
\label{fig:geometry_disksurface}
\end{figure*}

Near-infrared polarized scattered-light provides a direct geometric constraint on the height of the disk scattering-surface. For an inclined disk, the $\tau \sim 1$ scattering-surface projects to an ellipse whose center is offset from the stellar position along the minor axis \citep{de_boer_multiple_2016, ginski_direct_2016, avenhaus_disks_2018}. Under the assumption of an intrinsically circular and azimuthally symmetric disk, this offset can be converted into the scattering-surface height above the midplane $H_{\rm s}(r)$. We employ this geometry to infer the radial shape of $H_{\rm s}(r)$ and the corresponding flaring angle $\varphi(r)$ that regulates stellar irradiation, by utilizing the $r^{2}$ scaled scattered-light image (see Section~\ref{sec:results_nir_mm}).

The moderate inclination of V1094~Sco ($i = 54\arcdeg.7$) produces a measurable minor axis offset. In the $r^{2}$ scaled scattered-light image, the surface brightness can show two well-defined edges associated with the inner bright core at $r \sim 80$ au and the onset of the outer disk at $r \sim 200$ au. We consider that these edges mark the projected location of the $\tau \sim 1$ scattering-surface at two representative radii; they provide two anchor points for $H_{\rm s}(r)$.

We fit the two brightness edges with ellipses following \citet{Yamaguchi_alma_2025ApJ}. The edge positions used as constraints on the ellipse fits (white circles in Figure~\ref{fig:geometry_disksurface}) are extracted as follows. We first apply a Sobel filter implemented in $\tt scikit-image$ \citep{Walt2014PeerJ} to the $r^{2}$-scaled $Q_{\Phi}$ image to obtain a map of the local intensity gradient magnitude $|\nabla I|$, which highlights edge-like features. At each azimuthal angle, we then identify the edge as the radial position where $|\nabla I|$ reaches its local maximum. The resulting radial positions are azimuthally averaged within bins of $\Delta\theta \simeq 10\arcdeg$ width to suppress pixel-to-pixel noise, and these averaged positions are used as the data points for the ellipse fitting. Each ellipse is parameterized by its center $(x_{\rm c}, y_{\rm c})$, semi-major axis, semi-minor axis, and position angle. We fit only the near-side (northern) edge because the far-side (southern) edge is substantially fainter owing to back-scattering, and including it would increase the uncertainty in the inferred geometry \citep[see Figure~2 of][]{takami_surface_2014}. The best-fit ellipses are shown in Figure~\ref{fig:geometry_disksurface}, and the fitted parameters are listed in Table~\ref{tab:ellips_fit}.

We then convert the fitted ellipse centers into scattering-surface heights using the minor axis offset geometry. Defining the projected offset between the ellipse center and the stellar position as $O_{\rm c} = \sqrt{x_{\rm c}^{2} + y_{\rm c}^{2}}$, the scattering-surface height at the corresponding edge radius is
\begin{equation}
\label{eq:trigonometric_relation}
H_{\rm s}(r_{\rm s}) = \frac{O_{\rm c}}{\sin i}.
\end{equation}
The resulting aspect ratios are $H_{\rm s}/r = 0.067 \pm 0.040$ at the inner edge and $0.072 \pm 0.050$ at the outer edge (Table~\ref{tab:ellips_fit}). Within the uncertainties, these values indicate a shallow scattering-surface, consistent with substantial settling of small grains in the outer disk \citep{dullemond_effect_2004}.

With $H_{\rm s}$ measured at two radii, we describe the radial surface shape with a power law,
\begin{equation}
\label{eq:sca_surface_height}
\frac{H_{\rm s}(r)}{r} = \left(\frac{H_{0}}{r_{0}}\right)\left(\frac{r}{r_{0}}\right)^{\beta - 1},
\end{equation}
and constrain $(H_{0}, \beta)$ at $r_{0}=1~\mathrm{au}$ using an MCMC analysis with \texttt{emcee}. We adopt uniform priors and run 100 walkers for 500 steps, discarding the first 20 steps as burn in. We obtain $H_{0} = 0.05 \pm 0.01~\mathrm{au}$ and $\beta = 1.09 \pm 0.07$, consistent with passively irradiated flared disks \citep[$\beta \simeq 1.1$ to $1.3$;][]{kenyon_spectral_1987}.

Finally, we translate the inferred surface shape into the flaring (or grazing) angle that sets the intercepted stellar flux. Following \citet{Yamaguchi_alma_2025ApJ} and the irradiated disk models of \citet{kusaka_growth_1970} and \citet{chiang_spectral_1997}, the flaring angle is
\begin{align}
\varphi(r)
&= \frac{4}{3\pi}\frac{R_{\star}}{r}
   + r \frac{d}{dr}\left(\frac{H_{\rm s}(r)}{r}\right) \\
&\simeq r \frac{d}{dr}\left(\frac{H_{\rm s}(r)}{r}\right), \nonumber
\end{align}
where the finite stellar size term is negligible at the radii probed here. Substituting Equation~\ref{eq:sca_surface_height} yields
\begin{equation}
\varphi(r) = (0.004 \pm 0.003)
\left(\frac{r}{1~\mathrm{au}}\right)^{0.09 \pm 0.07}.
\end{equation}
At the outer edge of the scattered disk ($r_{\rm s} = 204~\mathrm{au}$), this corresponds to $\varphi \sim 0.01$, smaller than the $\varphi \sim 0.05$ often adopted for classical irradiated disks \citep{chiang_spectral_1997}. 

\begin{table}[h]
\begin{center}
\caption{Best Fit parameters of the elliptical fitting \label{tab:ellips_fit}}
\begin{tabular}{c@{\hskip 1mm}c@{\hskip 1mm}c@{\hskip 1mm}c@{\hskip 1mm}c}
\toprule
Location & \makebox[5mm][c]{PA} & \makebox[7mm][c]{$O_c$} & \makebox[13mm][c]{$r_{\rm s}$} & \makebox[13mm][c]{$H_{\rm s}$}\\
         & (deg)                & (au)                    & (au)                               & (au)\\
(1)      & (2)                  & (3)                     & (4)                                & (5)\\
\midrule
Inner disk & $108.4 \pm 1.9$ & $4.3 \pm 2.5 $ & $79.2 \pm 2.5$ & $5.3 \pm 3.1$ \\
Outer disk & $107.4 \pm 1.9 $ & $12.0 \pm 8.4$ & $203.5 \pm 3.6$ & $14.7 \pm 10.2$ \\
\bottomrule
\end{tabular}
\end{center}
\tablecomments{
Column descriptions: 
(1) Disk region adopted for the elliptical fitting. 
(2) Position angle of the fitted ellipse. 
(3) Geometric offset between the ellipse center and the stellar position. 
(4) Edge radius of the scattering-surface, corresponding to the semi major axis of the fitted ellipse. 
(5) scattering-surface height derived from the trigonometric relation in Equation~\ref{eq:trigonometric_relation}.}
\end{table}

\subsection{Disk Temperature}\label{sec:disktemp}

This section derives an observation-anchored estimate of the disk temperature profile by combining the scattering-surface geometry (Section~\ref{sec:shape_disksurface}) with a simple irradiation balance. The key input is the flaring angle $\varphi(r)$ inferred from the scattered-light morphology, which sets the fraction of stellar luminosity intercepted by the disk surface.

We adopt the framework of a passively heated, irradiated disk without dust scattering \citep{chiang_spectral_1997, okuzumi_global_2022}, in which the flaring angle $\varphi(r)$ regulates the fraction of stellar luminosity intercepted by the disk surface. The absorbed flux is expressed as $F_{\rm in} = \varphi(r)\,L_{\star}/8\pi r^{2}$ and is balanced by thermal re-emission from dust grains, $F_{\rm out} = \sigma_{\rm SB}\,T(r)^{4}$. Equating the two terms yields the radial temperature profile
\begin{align}\label{eq:disk_temp}
T_d(r)
&= \left[ \frac{L_{\star}\,\varphi(r)}{8\pi r^{2}\sigma_{\rm SB}} \right]^{1/4} \nonumber \\
&= (75 \pm 16)
  \left(\frac{r}{1~\mathrm{au}}\right)^{-0.48\pm0.02},
\end{align}
where $\sigma_{\rm SB}$ is the Stefan--Boltzmann constant and $L_{*} (= 0.64\pm0.14~L_{\odot})$ is the stellar luminosity derived from the Bayesian SED analysis using $\tt SHIDARE$, and provide the full methodology in Appendix~\ref{apx:stellar_properties}.

\begin{figure}[ht]
\centering
\includegraphics[width=0.46\textwidth]{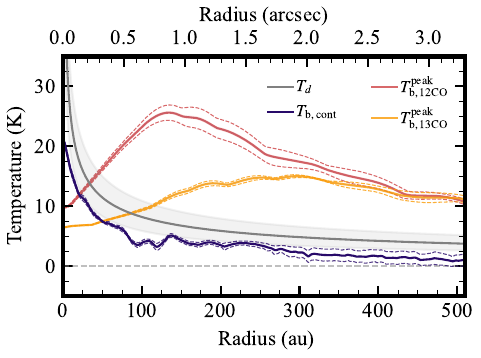}
\caption{Radial profiles of disk temperature in V1094~Sco. The solid gray curve shows the disk temperature $T_{\rm d}(r)$, with the shaded region indicating the $1\sigma$ uncertainty. The purple curve shows the dust continuum brightness temperature $T_{\rm b,cont}$, while the red and orange curves denote the peak brightness temperatures $T_{\rm b,12CO}^{\rm peak}$ and $T_{\rm b,13CO}^{\rm peak}$, respectively, derived from the moment--8 maps and converted using the Planck function. The azimuthal averaging procedure is identical to that used in Figure~\ref{fig:disk_radius}.}
\label{fig:disktemperature}
\end{figure}

As shown in Figure \ref{fig:disktemperature}, the derived temperature profile $T_{\rm d}(r)$ is consistently higher than the dust continuum brightness temperature, $T_{\rm b,cont}$, at all radii, as expected from radiative transfer. Over most of the disk radii, $T_{\rm d}(r)$ also remains below the CO peak brightness temperatures from both $^{12}$CO and $^{13}$CO, $T_{\rm b,CO}^{\rm peak}$, consistent with a vertically stratified structure in which optically thick CO traces warmer molecular layers above the midplane \citep{law_molecules_2021}. $T_{\rm b,CO}^{\rm peak}$ are computed from moment-8 maps derived from the same Keplerian-masked cubes used for the moment 0 and 1 maps. The masking suppresses noise and foreground contamination but does not modify the peak brightness of detected emission.

In the innermost region ($r\lesssim 100~\mathrm{au}$), $T_{\rm b,CO}^{\rm peak}$ declines toward the star. We verified that this trend is preserved when the peak brightness is measured from cubes imaged without continuum subtraction, indicating that it is not an artifact of the $\texttt{uvcontsub}$ procedure. A substantial fraction of this apparent suppression is likely observational. Because the synthesized beam ($0\farcs6 \simeq 90~\mathrm{au}$) subtends a large Keplerian velocity gradient, beam averaging and kinematic broadening dilute the peak line intensity, reducing the inferred $T_{\rm b,CO}^{\rm peak}$ even if the intrinsic gas temperature does not decrease.

Finally, we define an area-weighted mean temperature over the radial extent where the scattering-surface is constrained,
\begin{equation}
\langle T_{\rm d}\rangle \equiv
\frac{\int_{r_{\rm in}}^{r_{\rm out}} T_{\rm d}(r)\,r\,dr}{\int_{r_{\rm in}}^{r_{\rm out}} r\,dr},
\qquad
(r_{\rm in},r_{\rm out})=(1,204)~\mathrm{au}.
\end{equation}
\noindent
This definition is intentionally geometric and therefore depends on the adopted outer radius. For a power law $T_{\rm d}\propto r^{-\beta}$, the above mean is of order the outer edge temperature, $\langle T_{\rm d}\rangle \simeq [2/(2-\beta)]\,T_{\rm d}(r_{\rm out})$. With $\beta\simeq0.48$, this factor is $\simeq1.3$, implying that $\langle T_{\rm d}\rangle$ mainly reflects the temperature near $r_{\rm out}$. Using the outer edge of the scattered disk $r_{\rm out}=204~\mathrm{au}$ yields $\langle T_{\rm d}\rangle=8\pm2~\mathrm{K}$. Rather than interpreting this value as a universal disk averaged temperature, we regard it as a summary of the cold outer disk implied by the shallow flaring geometry.

\subsection{dust-trapping Diagnostics from Ring Widths}
\label{sec:dust_trapping}

To assess whether the observed rings are consistent with pressure-supported confinement of solids, we adopt the local gas pressure scale height $H_{\rm p}$ as a reference length scale for radial trapping. Using the irradiation-anchored temperature profile $T_{\rm d}(r)$ derived in Section~\ref{sec:disktemp}, we estimate $H_{\rm p}$ under vertical hydrostatic equilibrium as
\begin{equation}
\label{eq:scale_height}
H_{\rm p}(r)=\frac{c_s}{\Omega_{\rm K}}
= \sqrt{\frac{k_{\rm B}\,T_{\rm d}(r)\,r^{3}}{\mu\,m_{\rm p}\,G\,M_\star}},
\end{equation}
where $c_s=\sqrt{k_{\rm B}T_{\rm d}/(\mu m_{\rm p})}$ is the isothermal sound speed and $\Omega_{\rm K}=\sqrt{GM_\star/r^{3}}$ is the Keplerian angular frequency. We adopt $\mu=2.3$ and $M_\star=0.88\pm0.09~M_\odot$ (Section~\ref{sec:rotation_curve}), which yields
\begin{equation}
H_{\rm p}(r) = (0.019\pm0.002)\left(\frac{r}{1~\mathrm{au}}\right)^{1.26\pm0.01}~\mathrm{au}.
\end{equation}

Over the radii where the scattering-surface is constrained, the measured surface height $H_{\rm s}(r)=(0.05\pm0.01)(r/1~\mathrm{au})^{1.09\pm0.07}~\mathrm{au}$ (Section~\ref{sec:shape_disksurface}) corresponds to a few pressure scale heights, i.e., $H_{\rm s}(r)=\chi(r)\,H_{\rm p}(r)$ with $\chi$ of order unity to a few that may vary weakly with radius. This is broadly consistent with scattered-light radiative-transfer calculations in which the observed surface traces an optical-depth-unity layer located above the midplane when micron-sized grains remain well coupled to the gas \citep[e.g.,][]{muto_structure_2011, dong_missing_2012}. We therefore treat $H_{\rm p}$ as a physically motivated reference length scale for interpreting the radial confinement of the millimeter rings.

In gas-rich disks, radial drift of solids is driven by the pressure gradient and formally vanishes at local pressure maxima where $dP/dr=0$ \citep[e.g.,][]{takeuchi_radial_2002}. Such maxima act as convergence points for drifting particles and can sustain long-lived dust enhancements \citep[e.g.,][]{Muto2015PASJ, tsukagoshi_flared_2019, yen_kinematical_2020, liu_forming_2024}. A basic expectation of pressure-trap scenarios is that the characteristic radial width of a trapped ring should be comparable to, or smaller than, the gas pressure-support scale, because the width of a localized pressure perturbation is regulated by the gas rather than by the dust \citep[e.g.,][]{dullemond_disk_2018}. This criterion is intended as a consistency check: $\hat{w}_{\rm d} \lesssim H_{\rm p}$ is not unique to trapping, but it is a necessary condition for long-lived confinement by a relatively localized pressure structure.

We quantify intrinsic ring widths using the deconvolved Gaussian widths $\hat{w}_{\rm d}$ measured from the profile decomposition (Section~\ref{sec:results_substructures}) and evaluate $\hat{w}_{\rm d}/H_{\rm p}$ for the four well-defined rings at 107, 138, 187, and 231~au. The resulting ratios are $\hat{w}_{\rm d}/H_{\rm p}=1.1\pm0.1$, $1.0\pm0.1$, $0.6\pm0.1$, and $1.4\pm0.2$, respectively. The three inner rings have $\hat{w}_{\rm d}/H_{\rm p}\lesssim 1$ within the uncertainties and thus satisfy the necessary condition for confinement by a localized pressure perturbation. The outermost ring at 231~au has a slightly larger nominal ratio of $\simeq 1.4$, comparable to but somewhat exceeding the local pressure scale height; this value is most reasonably interpreted as broadly consistent with pressure-supported confinement, although less tightly confined than the inner three rings. Taken together, these widths motivate the dust-trapping interpretation developed further in Sections~\ref{sec:alpha_constraint} and \ref{sec:ring-gap_origin}, where the formation mechanism of individual rings is examined using additional diagnostics.

\section{Discussion}\label{sec:discussion}

\begin{figure*}[ht]
\centering
\includegraphics[width=0.98 \textwidth]{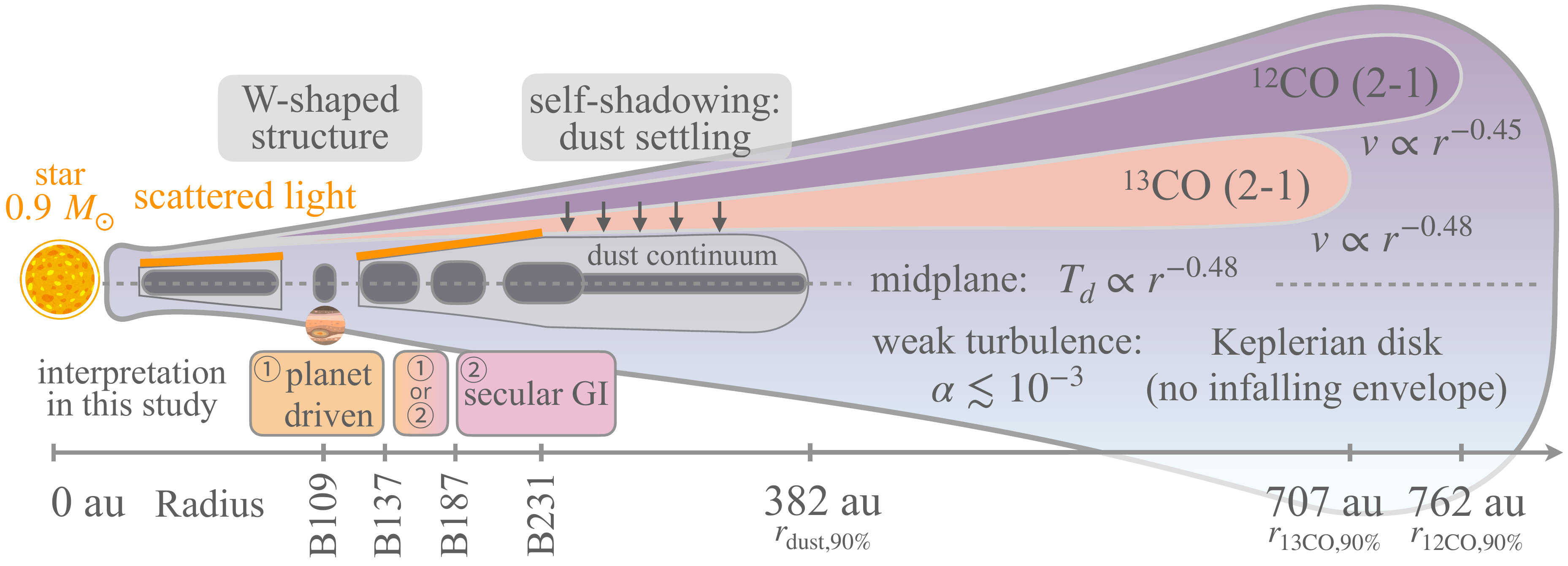}
\caption{Schematic cross section of the V1094~Sco disk summarizing the radial hierarchy of substructures and the emitting layers traced at different wavelengths. Dark gray annuli mark the prominent Band~6 dust continuum rings, including the W-shaped structure and its scattered-light counterpart. The orange band indicates the extent of the near-infrared scattered-light surface, which becomes fainter at large radii, consistent with a vertically settled outer disk and possible self-shadowing. The shaded regions illustrate the approximate emitting layers of the $^{13}$CO and $^{12}$CO lines. Vertical markers denote the radii enclosing $90\%$ of the integrated emission (curve ofgrowth) for the dust continuum and CO lines. The labels in the lower panel summarize the working interpretation adopted in this work: the intermediate W-shaped region is consistent with a planet-driven multigap response, whereas secular GI is considered as a competing interpretation for the outer ring--gap pairs.}\label{fig:schematic_view}
\end{figure*}

Figure~\ref{fig:schematic_view} provides a roadmap for interpreting the substructures in the exceptionally extended V1094~Sco disk. Our discussion is organized around three observational inferences: the rings are narrow (implying weak turbulent diffusion), the intermediate ``W''-shaped feature has a scattered-light counterpart (linking midplane and surface perturbations), and the two outer ring--gap pairs are regular yet lack scattered-light signatures (pointing to a midplane-dominated mechanism).

We proceed as follows. In Section~\ref{sec:alpha_constraint} we use the intrinsic ring widths to place conservative upper limits on the turbulent viscosity, finding $\alpha\lesssim10^{-3}$ across the outer disk. In Section~\ref{sec:ring-gap_origin} we test planet--disk interaction scenarios: the ensemble of gap widths and depths is inconsistent with a simple one-planet-per-gap picture (Section~\ref{sec:discussion_single_planet}), while the ``W''-shaped structure at $r\sim100$~au is consistent with multigap excitation by a single low-mass companion in a low-viscosity disk (Section~\ref{sec:discussion_lowmass_planet}). We then examine the two outermost ring-gap pairs, treating secular GI as a competing interpretation that naturally concentrates structure toward the midplane (Section~\ref{sec:SGI}). Finally, in Section~\ref{sec:SGI_planet} we discuss how large-scale solid concentration in a cold, weakly turbulent disk may facilitate planet formation at $\gtrsim100$~au, yielding a hybrid picture in which planet-driven and midplane-instability processes can operate at different radii.

\subsection{Turbulent viscosity from ring widths}
\label{sec:alpha_constraint}

In Section~\ref{sec:results_nir_mm}, we presented that the millimeter continuum extends to larger radii than the near-infrared scattered-light, a qualitative signature consistent with a vertically settled and weakly mixed outer disk. If turbulent stirring is indeed weak, it should also limit radial dust diffusion. We therefore use the intrinsic widths of the millimeter rings to place conservative upper limits on the turbulent viscosity parameter $\alpha$.

Turbulent diffusion acts as a stochastic broadening process for any dust concentration, whereas a pressure maximum can halt systematic radial drift but does not erase the diffusive random walk \citep[e.g.,][]{takeuchi_radial_2002}. Therefore, independent of the detailed ring formation pathway, the observed ring width provides an upper bound on the cumulative dust diffusivity over the time since the ring became confined.

To isolate this effect, we approximate the local evolution of the dust surface density perturbation by a linear diffusion equation,
\begin{equation}
\frac{\partial \Sigma_{\rm d}}{\partial t} = D_{\rm d}\,\frac{\partial^2 \Sigma_{\rm d}}{\partial r^2},
\end{equation}
for which a Gaussian profile remains self-similar and its variance evolves as
$w_{\rm d}^2(t)=w_{\rm d,0}^2+2D_{\rm d}t$.
Taking the most conservative limit $w_{\rm d,0}\rightarrow0$, the deconvolved width $\hat{w}_{\rm d}$ implies
\begin{equation}
D_{\rm d}\ \le\ D_{\rm d,obs}\equiv\frac{\hat{w}_{\rm d}^2}{2t_{\rm ring}},
\label{eq:upper_bound_d}
\end{equation}
where $t_{\rm ring}$ is the effective time available for diffusion after the ring became trapped.

We relate $D_{\rm d}$ to turbulent transport via $D_{\rm d}=(\nu_{\rm turb}/\mathrm{Sc})/(1+{\rm St}^2)$ \citep{youdin_particle_2007}, where $\mathrm{Sc}$ is the Schmidt number describing the ratio between turbulent viscosity and particle diffusivity. The turbulent viscosity is written as $\nu_{\rm turb}=\alpha c_s^2/\Omega_{\rm K}$ \citep{Shakura1973}. Assuming ${\rm St}\rightarrow0$ and $\mathrm{Sc}=1$, a commonly adopted value for well coupled particles, we obtain an indicative upper bound on $\alpha$ by equating $D_{\rm d}$ with the observed diffusivity $D_{\rm d,obs}$.
\begin{equation}
\alpha(r_0)\ \le\ \frac{1}{2\,\Omega_{\rm K}(r_0)\,t_{\rm ring}}\left(\frac{\hat{w}_{\rm d}}{H_{\rm p}(r_0)}\right)^2 .
\end{equation}
\noindent
This expression means that the inferred turbulent viscosity depends on the dimensionless ring width $(\hat{w}_{\rm d}/H_{\rm p})$, with narrower rings implying smaller values of $\alpha$. Using the dynamical stellar mass $M_\star$ (Section~\ref{sec:rotation_curve}) and the disk temperature profile
$T_{\rm d}(r)$ (Section~\ref{sec:disktemp}),
we obtain
\begin{align}
\alpha(r_{0})
&\le
(2.2\pm0.5)\times10^{-3}
\left(\frac{\hat{w}_{\rm d}}{10\,\mathrm{au}}\right)^{2}
\left(\frac{t_{\rm ring}}{10^5\,\mathrm{yr}}\right)^{-1}
\nonumber\\
&\hspace{7em}\times
\left(\frac{r_{0}}{100\,\mathrm{au}}\right)^{-1.02\pm0.02}.
\label{eq:alpha_scaling_result_final}
\end{align}
We adopt $t_{\rm ring}=10^{5}$--$10^{6}$~yr. The upper end is bounded by the stellar age ($\simeq2.4\times 10^{6}$ yr in Appendix \ref{apx:stellar_properties}), while the lower bound corresponds to the time required for turbulent diffusion to broaden a ring over a scale comparable to the observed width, $t_{\rm diff}\sim \hat{w}_{\rm d}^{2}/(2\alpha c_s H_{\rm p})$. For $\alpha\sim10^{-3}$, this timescale is of order \(10^{5}\)~yr at \(r\sim100\)–\(200\)~au in the cold outer disk.
For weaker turbulence, \(\alpha\sim10^{-4}\), the diffusion time increases to \(\sim10^{6}\)~yr, comparable to the stellar age. Thus, adopting \(t_{\rm ring}\gtrsim10^{5}\)~yr represents a conservative choice that avoids artificially tightening the upper limit on \(\alpha\). Shorter lifetimes would imply that the rings are observed before turbulent diffusion can substantially modify their widths, suggesting that the present morphology would have to correspond to a relatively early evolutionary stage.

Applying Equation~(\ref{eq:alpha_scaling_result_final}) to the four well-defined outer rings (Table~\ref{tab:prop_disk_substructures}) yields $\alpha\lesssim10^{-3}$ for $t_{\rm ring}=10^{5}$~yr and $\alpha\lesssim10^{-4}$ for $t_{\rm ring}=10^{6}$~yr. An independent constraint supports this inference. \citet{villenave_turbulence_2025} report a complementary upper limit on the turbulence parameter in V1094~Sco (denoted $\alpha_{\rm frag}$, derived from the vertical settling constraint under a fragmentation-limited grain-growth assumption), $\alpha_{\rm frag} \lesssim 1.9 \times 10^{-3}$ at the outer ring ($200$--$280$~au). Although they constrain distinct components of the turbulence (vertical versus radial), their upper limit is consistent with our radial-diffusion-based bound.

\subsection{Origin of the Ring–gap Structures}\label{sec:ring-gap_origin}

In this subsection, we examine the origin of the ring–gap structures by comparing their observed properties with predictions from theoretical models.

\subsubsection{Discrepancy of a Planet Opening a Single Ring--Gap Pair}
\label{sec:discussion_single_planet}

First, we compare the observed gaps in V1094~Sco with the planet--disk interaction framework of \citet{zhang_disk_2018}, who performed two-dimensional hydrodynamic simulations that include both gas and dust and derived a predictive relation between the gap depth, $\delta_{\rm I}$, and the normalized gap width, $\Delta_{\rm I}$, measured from the surface-brightness profile $I_{\nu}(r)$. Using Equations~(22)--(24) of \citet{zhang_disk_2018}, the relationship can be written as
\begin{align}
\Delta_{\rm I}
= A \left[
0.635
\left(\frac{H_{\rm p,gap}}{r_{\rm gap}}\right)^{2.63}
\left(\frac{\alpha}{10^{-3}}\right)^{0.07}
\left(\frac{\delta_{\rm I}-1}{C}\right)^{1/D}
\right]^B ,
\label{eq:width_depth}
\end{align}
where $H_{\rm p,gap}/r_{\rm gap}$ is the disk aspect ratio at the gap location and $\alpha$ is the viscosity parameter. The coefficients $A$, $B$, $C$, and $D$ depend on the gas surface density, $\Sigma_{\rm gas}$, and grain properties. We note that the planet--star mass ratio $q$ is eliminated from Equation~\ref{eq:width_depth} by combining the two empirical relations of \citet{zhang_disk_2018}, which relate $q$ separately to $\delta_{\rm I}$ and to $\Delta_{\rm I}$. The resulting predictive locus in the $(\Delta_{\rm I}, \delta_{\rm I})$ plane therefore depends on the disk properties but not on the planet mass itself. We emphasize that this relation is derived under the assumption that a single planet on a circular orbit opens a single, isolated gap-ring pair in a steady state, which constitutes the working hypothesis tested in this subsection.

To bracket plausible disk conditions, we consider a wide range of parameters, adopting $\Sigma_{\rm gas}=1\sim100~\mathrm{g~cm^{-2}}$ and the aspect ratio over $H_{\rm p,gap}/r_{\rm gap}=0.05\sim0.1$. The aspect ratio inferred from the temperature profile of V1094~Sco is $H_{\rm p}/r \approx 0.06$--$0.08$ over the radii of the outer rings (see Section~\ref{sec:dust_trapping}). To account for possible uncertainties, we therefore explore a slightly broader range of $H_{\rm p,gap}/r_{\rm gap}$ in the comparison with the planet–disk interaction models.

\begin{figure}[ht!]
\centering
\includegraphics[width=0.45 \textwidth]{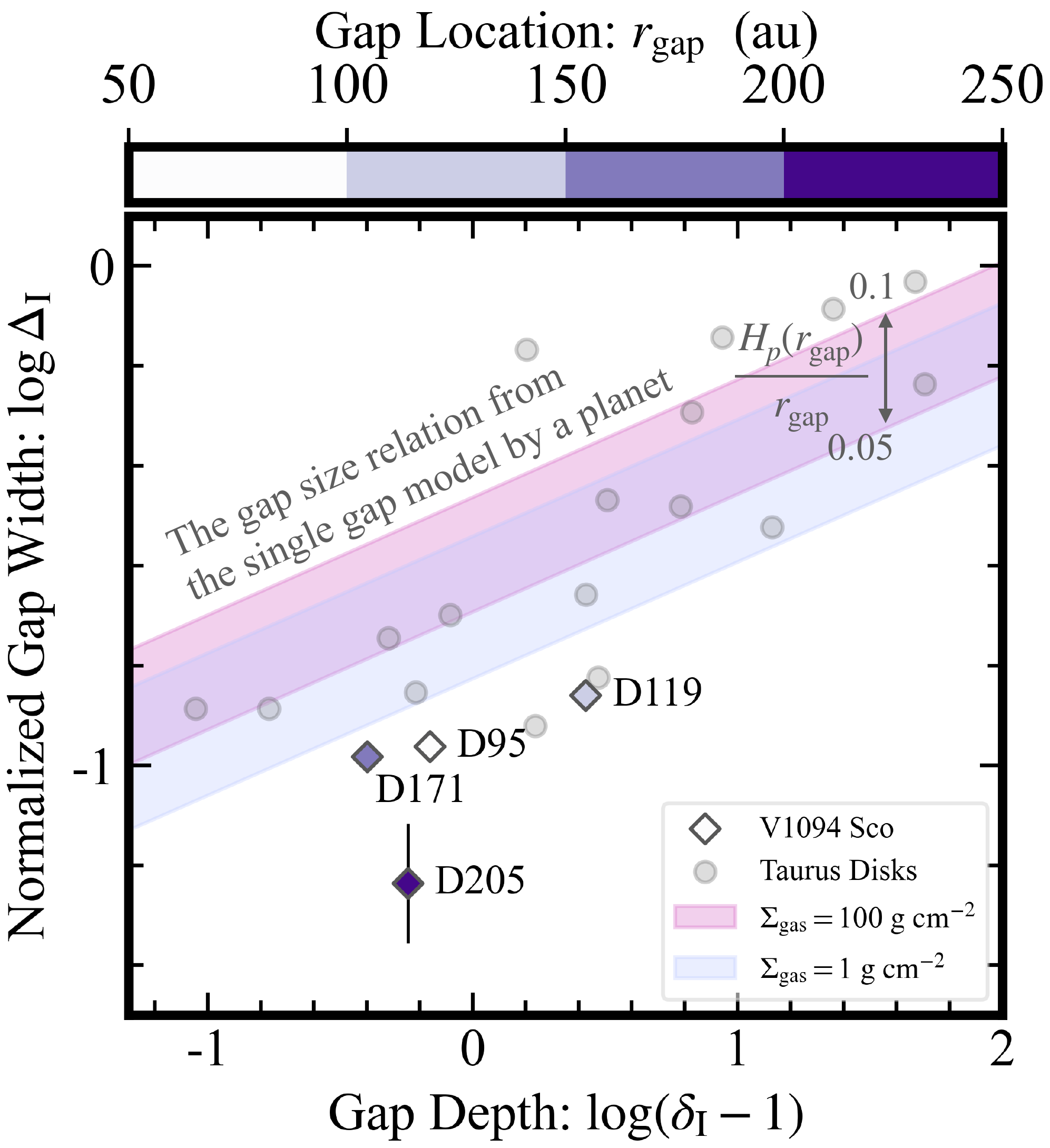}
\caption{Comparison between the scaling relation of gap width and depth predicted by the single-planet gap model of \citet{zhang_disk_2018} (shaded regions) and the spatially resolved gaps observed in the V1094~Sco disk (diamond symbols). For reference, gaps measured in Taurus disks from \citet{yamaguchi_alma_2024} are shown as circles. The models assume a fixed maximum dust grain size of 0.1~mm and a viscous parameter of $\alpha=10^{-3}$. Model regions are shown for gas surface densities of $\Sigma_{\rm g}=100~\mathrm{g\,cm^{-2}}$ (purple) and $1~\mathrm{g\,cm^{-2}}$ (blue). 
For each surface density, the shaded regions span a range of disk scale heights from $H_{p}(r_{\rm gap})/r_{\rm gap}=0.05$ to $0.1$.}
\label{fig:comparison_planetmodel}
\end{figure}

We employ the DSD1 model of \citealt{zhang_disk_2018} which follows dust sizes as a power-law distribution ranging from $0.005~\mu \rm m$ to 0.1 mm with a power-law index of $-3.5$, and we adopt $\alpha=10^{-3}$. The resulting model expectations are shown as shaded regions in Figure~\ref{fig:comparison_planetmodel}.

All spatially resolved gaps in V1094~Sco fall outside these predicted regions, exhibiting systematically narrower normalized widths than expected for gaps opened by a single planet under the explored parameter space. This behavior contrasts with the Taurus sample studied by \citet{yamaguchi_alma_2024}, where the majority of Class~II disks show gap properties broadly consistent with the \citet{zhang_disk_2018} predictions (see circle markers in Figure~\ref{fig:comparison_planetmodel}). We have further verified that adopting a larger maximum grain size (e.g., DSD2 model of \citealt{zhang_disk_2018} with $a_{\rm max} = 1$~cm) does not alter this finding because the observed gaps remain outside the predicted single-planet locus across the explored $\Sigma_{\rm gas}$ range. The persistence of the discrepancy in V1094~Sco, even after allowing for a wide range of $\Sigma_{\rm gas}$, $H_{\rm p,gap}/r_{\rm gap}$, and dust grain size distribution, suggests that a simple scenario in which each observed gap is produced by an isolated, single planet carving a single ring--gap pair is unlikely to be the dominant explanation for this system.

\subsubsection{A low-mass Planet as the Origin of the W-shaped Structure}\label{sec:discussion_lowmass_planet}

The systematic failure of the single-planet and single-gap framework (Section~\ref{sec:discussion_single_planet}) implies that the substructures in V1094~Sco are unlikely to be explained as a straightforward superposition of independent ring-gap pairs carved by multiple isolated planets. An alternative interpretation is that at least part of the architecture reflects a coherent disk response to wave excitation by a dominant companion in a weakly turbulent disk, rather than a one-to-one correspondence between individual planets and individual gaps.

Several observational properties of the millimeter continuum favor a common dynamical origin for multiple features. Most prominently, the W-shaped morphology formed by two adjacent gaps (D95 and D119) bracketing a bright ring (B109) closely resembles the multigap patterns produced in low-viscosity disks by a single embedded companion through secondary and higher-order gap formation \citep[e.g.,][]{bae_formation_2017, dong_multiple_2017, dong_multiple_2018, kanagawa_model_2020, kuwahara_dust_2022, kuwahara_dust_2024}. The presence of a scattered-light depression at the same stellocentric radius as this W-shaped structure (Section~\ref{sec:results_nir_mm}) further indicates that the underlying perturbation affects both the midplane and the disk's surface, as expected for a planet induced disturbance in a weakly turbulent disk \citep[e.g.,][]{dong_multiple_2017}. In addition, the mild curvature feature exterior to B137 (I159) is qualitatively consistent with size dependent dust transport in the vicinity of a pressure perturbation, although we treat this as a suggestive rather than decisive indicator \citep{bi_shoulder_2024}.

These clues motivate a regime in which a single low-mass companion excites spiral density waves that propagate away from the planet and steepen into shocks, producing multiple dust gaps at finite radial separations. In this picture, decreasing turbulent diffusion enhances the contrast and longevity of secondary and higher-order gaps, allowing a single planet to generate a multigap morphology \citep{bae_formation_2017}. Importantly, existing simulations demonstrate that additional gaps can arise at radii well separated from the primary double-gap system, particularly in cold geometrically thin disks where wave propagation and dissipation can generate higher-order structures \citep[e.g.,][]{bae_formation_2017, bae_planet-driven_2018}. Thus, while the W-shaped structure provides the clearest anchor for a planet-driven interpretation, a planet origin cannot be excluded a priori for more distant, lower contrast substructures.

A useful mass scale for characterizing this interaction is the thermal mass,
\begin{equation}
M_{\rm th} \equiv \left(\frac{H_{\rm p}(r_p)}{r_{\rm p}}\right)^3 M_{\star},
\end{equation}
which sets the scale at which spiral density waves shock close to the planet orbit \citep{Goodman2001ApJ}. For $M_{\rm p}\lesssim M_{\rm th}$, waves can travel a finite radial distance before shocking, allowing a single companion to generate multiple gaps at distinct radii.

Following \citet{dong_multiple_2018}, we estimate the companion mass from the separation of the double gaps using their empirical relation (their Equation~11),
\begin{equation}
\frac{M_{\rm p}}{M_{\oplus}} =
\frac{1.144 \times 10^7}{\gamma+1}
\left(\frac{M_{\star}}{M_{\odot}}\right)
\left(\frac{H_{\rm p}(r_{\rm p})}{r_{\rm p}}\right)^{11/2}
\left(\frac{r_{\rm p}}{\Delta_{\rm gap}}\right)^{5/2},
\end{equation}
where we adopt $\gamma=1$ for isothermal gas and $\Delta_{\rm gap}$ is the radial separation between the two gaps.
For the W-shaped structure, we take $\Delta_{\rm gap}=D119-D95=24$~au and place the companion at $r_{\rm p}=109$~au (B109).
Using the dynamical stellar mass $M_{\star}$ from Section~\ref{sec:rotation_curve} and the pressure scale height $H_{\rm p}(r)$ from Section~\ref{sec:dust_trapping}, we obtain $M_{\rm p} = (0.8 \pm 0.6)\,M_{\rm th} = (55 \pm 35)\,M_{\oplus}$.
Because of the steep dependence on $H_{\rm p}/r$, the absolute mass scale remains sensitive to systematic uncertainties in the thermal and vertical disk structure, but the inferred order unity value of $m\equiv M_{\rm p}/M_{\rm th}$ supports a companion near the thermal mass scale.

We further test whether the W-shaped morphology is compatible with the type of gas flow expected around such a companion, following \citet{kuwahara_influences_2020-1} and \citet{kuwahara_dust_2022}. This is not an independent planet detection but a consistency check on the flow regime required to sustain the observed morphology. In their framework, double-gap (here, W-shape) dust structures induced from a planet arise in a flow shear regime rather than a flow headwind regime. Since the radial pressure gradient is not directly measured here, the condition for the flow shear regime can be written as
\begin{equation}
\left| \frac{d\ln p}{d\ln r} \right|
< \frac{15}{2}\,\frac{m}{h},
\end{equation}
with $h\equiv H_{\rm p}/r$. At $r_{\rm p}=109~\mathrm{au}$, our observation-anchored disk structure yields $m\simeq0.8$ and $h\simeq0.06$, corresponding to the weak requirement $\left|d\ln p/d\ln r\right|\lesssim100$. Since realistic disks have pressure gradients of order unity \citep[e.g.,][]{ida_radial_2016}, the inferred parameters place the system well within the flow shear regime.

Taken together, the W-shaped continuum structure, its scattered-light correspondence, and the near thermal estimate of $m$ support a coherent interpretation in which a single low-mass companion embedded in a weakly turbulent disk can drive multiple gaps through spiral wave excitation and dissipation. At the same time, the outer ring-gap pairs show additional phenomenology, including the absence of scattered-light counterparts and their apparent regularity, that may indicate contributions from other midplane-dominated processes. We therefore treat the planet-driven picture as a plausible but not exclusive explanation for the multiring architecture, and in the next subsection we examine secular GI as a competing mechanism for the outermost structures.

\subsubsection{Nature of the Outer Ring--Gap Pairs}\label{sec:outer_pairs}

We now examine the origin of the two outer ring--gap pairs, D171/B187 and D205/B231, located at the largest stellocentric distances in the V1094~Sco disk. The preceding sections show that the single-planet single-gap framework fails to reproduce the ensemble of gap widths and depths (Section~\ref{sec:discussion_single_planet}), while the W-shaped structure at $\sim100$~au is naturally explained by a low-mass companion through secondary and higher-order gap formation in a low-viscosity disk (Section~\ref{sec:discussion_lowmass_planet}). A key question is therefore whether the two outermost pairs represent a distant extension of the same planet-driven response or instead arise from a distinct mechanism operating in the outer disk.

The outer pairs exhibit two properties. First, they have similar widths and depths (Table~\ref{tab:disk_substrucutures}). Second, they lack clear scattered-light counterparts, unlike the W-shaped structure that coincides with a depression in the $r^{2}$-scaled near-infrared image (Section~\ref{sec:results_nir_mm}). While nondetections in scattered-light do not by themselves exclude a planetary origin, they favor mechanisms that concentrate contrast near the midplane without producing an equally strong response at the scattering-surface.

A planet-driven interpretation remains viable in principle. Numerical simulations have shown that secondary gap structures can emerge exterior to the primary gap system, particularly in cold, geometrically thin disks where spiral wave propagation and dissipation generate a broader hierarchy of substructures \citep[e.g.,][]{bae_formation_2017, bae_planet-driven_2018}. In this framework, the D171/B187 ring-gap pair could arise as a more distant response to the same companion that accounts for the W shaped morphology. By contrast, the present data do not require the more distant D205/B231 pair to share this origin. The current observations therefore do not rule out a planet driven origin for D171/B187.

Alternative explanations tied to disk chemistry are disfavored. Models in which variations in dust growth and fragmentation near snowlines generate rings and gaps generally predict prominent signatures in micron-sized grains and therefore in near-infrared scattered-light \citep{pinilla_dust_2017}. The lack of scattered-light counterparts at the locations of D171/B187 and D205/B231 therefore does not support a snowline-related origin for these outer structures.

Magnetically driven mechanisms provide another nonplanetary pathway. Simulations of wind-driven disks and nonideal magnetohydrodynamic disks show that radially structured magnetic stresses and wind mass loss can generate pressure perturbations and ring-like dust concentrations \citep{flock_gaps_2015}. Because these stresses can act over a substantial vertical extent, they may also perturb the surface layers and thereby leave signatures in small grains traced in scattered-light \citep[e.g.,][]{suriano_formation_2019}. The absence of clear scattered-light counterparts at D171/B187 and D205/B231 therefore makes an MHD-driven origin less likely unless the surface response is strongly suppressed by illumination geometry or grain-dependent effects.

These considerations motivate us to explore mechanisms that naturally operate in the cold and weakly turbulent outer disk.

\subsubsection{Secular Gravitational Instability as a Possible Origin}\label{sec:SGI}

In this context, secular GI provides a physically motivated alternative for the formation of the outer ring--gap pairs \citep{youdin_formation_2011, takahashi_two-component_2014}. Secular GI operates in dust-rich disks with weak turbulence and develops as a midplane-concentrated instability mediated by dust self-gravity and gas drag, while remaining comparatively ineffective in vertically extended gas layers \citep{tominaga_secular_2023}. This intrinsic vertical selectivity offers a natural pathway to producing pronounced millimeter continuum substructures without equally strong signatures in scattered-light.

Secular GI also tends to generate multiple relatively narrow rings with quasi-regular spacing at large radii, even when the disk is globally gravitationally stable \citep{takahashi_origin_2016}. A simple observational consistency check is therefore provided by the spacing between adjacent rings that bracket the outer gap pairs. Following \citet{tominaga_revised_2019}, secular GI is expected to operate efficiently when $\Delta r_{\rm ring}/H_{\rm p} < 4$. For the separations B137--B187 and B187--B231, which bracket the two outer ring pairs, we obtain $\Delta r_{\rm ring}/H_{\rm p} = 2.9 \pm 0.4$ and $4.1 \pm 0.5$, respectively. The former lies comfortably within the nominal secular GI efficient regime, while the latter is consistent with marginal operation near the boundary.

A more quantitative test can be performed using the secular GI criterion for vertically stratified disks derived by \citet{tominaga_secular_2023}. Rewriting their condition (their Equation~74) in terms of standard transport parameters, and approximating the radial dust diffusivity as $D_r \simeq \alpha\,c_s H_{\rm p}$, the secular GI condition can be expressed as

\begin{equation}
\label{eq:SGI_criterion}
\begin{aligned}
\frac{M_{\rm d,loc}(r)}{M_\odot}
\ \gtrsim\ 
&\,3\times10^{-4}
\left(\frac{M_\star}{M_\odot}\right)
\left(\frac{\Sigma_{\rm d}/\Sigma_{\rm g}}{10^{-2}}\right)^{1/2} \\
&\times
\left(\frac{H_{\rm p}/r}{10^{-1}}\right)
\left(\frac{\alpha/{\rm St}}{10^{-3}}\right)^{1/2}.
\end{aligned}
\end{equation}
Here, we define $M_{\rm d,loc}(r)\equiv \pi r^{2}\Sigma_{\rm d}(r)$ as a local characteristic dust mass scale. Although the quantities entering Equation~(\ref{eq:SGI_criterion}) are local values at a given radius, $M_{\rm d,loc}(r)$ roughly represents the enclosed dust disk mass when expressed in terms of local disk quantities \citep{tominaga_secular_2023}. Here, $\Sigma_{\rm d}$ and $\Sigma_{\rm g}$ denote the dust and gas surface densities, respectively. The numerical prefactor represents an order-of-magnitude estimate of the instability threshold under fiducial disk conditions and should be interpreted as an approximate criterion rather than a sharp boundary.

The radial dust surface density profile $\Sigma_{\rm d}(r)$ is derived directly from the azimuthally averaged dust continuum intensity profile of the PRIISM model image (Figure \ref{fig:priism_radialprofile}) using a radiative-transfer relation, $\Sigma_{\rm d}(r) = -\ln \left[1 - I_{\nu}(r)/B_{\nu}(T_{\rm d}(r))\right]/\kappa_{\nu}$, where $B_{\nu}$ is the Planck function, $T_{\rm d}(r)$ is the disk temperature profile derived in Section~\ref{sec:disktemp}, and $\kappa_{\nu}$ is dust absorption opacity, adopting fiducial $\kappa_{\nu}=2.3~\mathrm{cm}^{2}~\mathrm{g}^{-1}$ at 1.3 mm \citep{beckwith_particle_1991}.

To evaluate Equation~(\ref{eq:SGI_criterion}) for V1094~Sco, we adopt representative transport parameters for the outer disk. The Stokes number in the Epstein regime is ${\rm St}\simeq (\pi/2)\rho_{\rm s}s_{\rm d}/\Sigma_{\rm g}$, where $\rho_{\rm s}$ and $s_{\rm d}$ are the internal density and characteristic grain size. We adopt $\rho_{\rm s}=3~\mathrm{g~cm^{-3}}$ and a fiducial grain size $s_{\rm d}\sim1~\mathrm{mm}$\footnote[5]{While we adopt a fiducial large-grain size of $s_{\rm d}\sim1$~mm, polarization-based self-scattering studies suggest maximum grain sizes of $\lesssim100~\mu$m in some disks \citep{bacciotti_alma_2018, Liu2026_pol}. A smaller $s_{\rm d}$ reduces ${\rm St}$ linearly, but $\kappa_\nu$ likewise varies with the maximum grain size \citep{birnstiel_disk_2018}; because $\Sigma_{\rm d}\propto 1/\kappa_\nu$ for a fixed observed intensity, the inferred $\Sigma_{\rm g}$ (for fixed $\epsilon_{\rm dg}$) changes in tandem. Thus, ${\rm St}\propto s_{\rm d}\kappa_\nu$, so adopting $s_{\rm d}\sim100~\mu$m mainly shifts the quantitative value of $\alpha/{\rm St}$ without altering our qualitative conclusion.} \citep[e.g.,][]{kawasaki_impact_2025}. With $\epsilon_{\rm dg}\equiv\Sigma_{\rm d}/\Sigma_{\rm g}=10^{-2}$ and the measured $\Sigma_{\rm d}=0.1$--$0.2~\mathrm{g~cm^{-2}}$ at $r=100$--$230~\mathrm{au}$, we obtain ${\rm St}=(2$--$5)\times10^{-2}$. Independent constraints from the ring survival analysis (Section~\ref{sec:alpha_constraint}) suggest $\alpha=10^{-4}$--$10^{-3}$ in the same radial range. These ranges imply $\alpha/{\rm St}=(2\times10^{-3})$--$(4\times10^{-2})$ for the fiducial case. Rather than fixing a single value, we evaluate the secular GI criterion for two representative cases, $\alpha/{\rm St}=10^{-3}$ and $10^{-2}$, which bracket the weak turbulence regime implied by these constraints.

\begin{figure}[t]
\centering
\includegraphics[width=0.48 \textwidth]{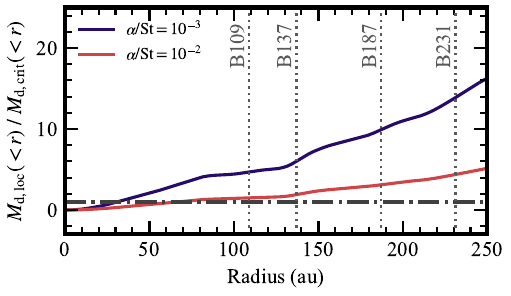}
\caption{Quantitative test of secular GI in the V1094~Sco disk. Radial profile of the ratio between the enclosed dust mass, $M_{\rm d, loc}(<r)$, and the critical dust mass for the onset of secular GI, $M_{\rm d,crit}(<r)$, evaluated using the vertically stratified secular GI criterion. Two representative values of the transport parameter $\alpha/{\rm St}$ are shown: $\alpha/{\rm St}=10^{-3}$ (blue curve) and $10^{-2}$ (red curve). Vertical dotted lines indicate the locations of the dust rings B109, B137, B187, and B231. The region where $M_{\rm d,loc}(r)/M_{\rm d,crit}(r)\gtrsim1$ corresponds to radial zones in which secular GI can plausibly operate, and the horizontal dashed line indicates its border.}\label{fig:secular GI_criterion}
\end{figure}

Figure~\ref{fig:secular GI_criterion} presents the radial dependence of the ratio $M_{\rm d,loc}(r)/M_{\rm d,crit}(r)$, where $M_{\rm d,crit}(<r)$ is defined by Equation~(\ref{eq:SGI_criterion}). This diagnostic evaluates the secular GI criterion for two cases, depending on $\alpha/{\rm St}=10^{-3}$ and $10^{-2}$. We emphasize that this diagnostic does not simply reflect the monotonic increase of enclosed dust mass with radius. Instead, the critical mass itself scales with radius through the pressure scale height and transport parameters, such that the ratio quantifies whether the local disk conditions at a given radius are favorable for secular GI. Regions where $M_{\rm d,loc}(r)/M_{\rm d,crit}(r)\gtrsim1$ therefore indicate radial zones in which secular GI can plausibly operate, rather than a trivial consequence of integrating more mass at larger radii. Under these conditions, the criterion can be satisfied not only near B187 and B231 but also near B109 and B137. This result suggests that secular GI is physically plausible over a wide radial extent of the disk and therefore constitutes a meaningful competing interpretation for the outer ring-gap pairs, alongside the planet-driven multigap scenario discussed above.

Throughout this analysis we adopt a fiducial dust-to-gas ratio of $\epsilon_{\rm dg}=10^{-2}$, which is commonly assumed for disks. If the local dust enrichment were higher (i.e., $\epsilon_{\rm dg}$ closer to $10^{-1}$), the inferred gas surface density would be lower for a fixed $\Sigma_{\rm d}$, leading to a larger Stokes number and hence a smaller $\alpha/{\rm St}$. In this sense, our choice of $\epsilon_{\rm dg}=10^{-2}$ does not bias the analysis toward triggering secular GI; rather, it represents a conservative assumption that tends to make the secular GI condition more difficult to satisfy.

We note that the dust surface density profile adopted here is derived under the assumption that the continuum emission is at most marginally optically thin. If instead the inner disk ($r\lesssim70$~au) is optically thick \citep{van_terwisga_v1094_2018}, and if dust scattering in such optically thick regions further suppresses the emergent continuum intensity \citep[e.g.,][]{liu_anomalously_2019}, then the dust mass in the inner regions would be systematically underestimated. In that case, the true enclosed dust mass $M_{\rm d,loc}(<r)$ would be higher than inferred here, leading the disk even more susceptible to secular GI. Our analysis therefore provides a conservative lower limit on the susceptibility to secular GI in the inner disk.

\subsection{Secular GI as a Cradle for Planet Formation}\label{sec:SGI_planet}

The discussion so far has focused on the dynamical origin of the ring--gap architecture and its possible connection to an embedded planet at stellocentric radii of $r\sim100$~au. A central question is how planet formation can proceed at such large distances, where classical growth timescales are expected to exceed disk lifetimes \citep[e.g.,][]{rafikov_fast_2004, ida_toward_2008}. In V1094~Sco, the pronounced ring morphology implies efficient redistribution and concentration of solids; the key issue is whether this disk-wide concentration can also facilitate the earliest stages of planet formation.

A key ingredient emerging from our analysis is the cold, weakly turbulent outer disk. For vertical hydrostatic equilibrium, the pressure scale height satisfies $H_{\rm p}=c_s/\Omega_{\rm K}\propto T_{\rm d}^{1/2}$ at fixed stellar mass and radius. A lower temperature therefore implies a smaller $H_{\rm p}$ and, for a given surface density, a higher midplane gas density, $\rho_{\rm g}\propto\Sigma_{\rm g}/H_{\rm p}$. Once dust is vertically settled, the midplane dust density follows an analogous scaling, $\rho_{\rm d}\propto\Sigma_{\rm d}/H_{\rm d}$, with $H_{\rm d}\lesssim H_{\rm p}$ in weak turbulence. In such environments, even modest surface density enhancements can translate into substantial increases in midplane dust concentration, creating conditions favorable for efficient solid growth.

Within this context, secular GI offers an appealing framework (see Section \ref{sec:SGI}), which is conceptually analogous to other collective dust instabilities such as the streaming instability \citep{youdin_streaming_2005, Johansen2007Natur}, but it operates preferentially on global ring forming scales and may precede later stage planetesimal assembly \citep{takahashi_origin_2016}. Rather than being a complete planet formation mechanism on its own, secular GI can be viewed as a process that organizes solids into a ring dominated architecture and enhances midplane dust concentrations, thereby providing favorable initial conditions for subsequent growth channels; this remains true even when solids undergo radial drift \citep{tominaga_secular_2020}.

The V1094~Sco disk further suggests that secular GI driven organization need not be confined to the outermost regions. The intermediate rings B109 and B137, associated with the W-shaped continuum morphology and the planet-based interpretation discussed in Section~\ref{sec:discussion_lowmass_planet}, also reside at radii where the secular GI condition can be satisfied. This raises the possibility that the planet responsible for the W-shaped structure did not form independently of the ring system, but instead represents a later outcome in a disk where secular GI had already organized solids into long-lived concentrations.

To examine whether secular GI remains compatible with the companion mass scale inferred at $r\simeq109$~au, we consider a simple mass budget argument. If a solid mass reservoir associated with a forming companion is redistributed into an annulus with radial width comparable to the local pressure scale height, $\Delta r\sim H_{\rm p}$ \citep{tominaga_secular_2020}, the implied dust surface density is $\Sigma_{\rm d}\sim M_{\rm solid}/(2\pi r H_{\rm p})$. The local mass scale entering Equation~(\ref{eq:SGI_criterion}) is then $M_{\rm d,loc}\sim \pi r^{2}\Sigma_{\rm d}\sim (r/2H_{\rm p})\,M_{\rm solid}$.

With $H_{\rm p}/r\simeq0.06$ at $r=109$~au (Section~\ref{sec:dust_trapping}) and $\alpha/{\rm St}\sim10^{-3}$–$10^{-2}$ (Section~\ref{sec:SGI}), Equation~(\ref{eq:SGI_criterion}) corresponds to a minimum solid reservoir of $M_{\rm solid}\simeq7$–$23\,M_\oplus$ for the fiducial dust-to-gas ratio ($\epsilon_{\rm dg}=10^{-2}$) adopted here. This estimate should be interpreted as an order-of-magnitude requirement for the solid mass available to assemble a planetary companion.

In Section~\ref{sec:discussion_lowmass_planet}, the W-shaped double gap at $r_{\rm p}\simeq109$~au yields a companion mass estimate of $M_{\rm p}=(55\pm35)\,M_\oplus$. This mass scale is comparable to, though somewhat larger than, the minimum solid reservoir estimated above. If secular GI operates once the local dust surface density exceeds the instability threshold, it may initially form a lower-mass planetary core that subsequently grows through continued accretion. While the present data do not constrain this growth history, the comparison suggests that the solid mass required for triggering secular GI is broadly consistent with the companion mass scale implied by the observed gaps.

Taken together, V1094~Sco appears to host a hybrid architecture enabled by weak turbulence. At intermediate radii, a low-mass companion can sculpt multiple gaps through spiral wave excitation and dissipation, while at larger radii a global instability concentrated toward the midplane can organize solids into narrow, quasi-regular rings (Figure~\ref{fig:schematic_view}). In this picture, secular GI can act as a cradle for planet formation by establishing dis-wide conditions under which planets can form efficiently and persist in an exceptionally cold and extended disk.

\section{Conclusion}\label{sec:conclusion}

We have conducted a comprehensive multiwavelength study of the protoplanetary disk around V1094~Sco, combining ALMA Band~6 continuum and CO isotopologue data with an archival VLT/SPHERE near-infrared scattered-light image. By uniformly reprocessing all available ALMA datasets and employing both super-resolution and restored imaging, we resolve the disk morphology, kinematics, and thermal structure over nearly three orders of magnitude in radius. Our main findings are summarized below.

\begin{enumerate}

\item{V1094~Sco hosts one of the most radially extended Class~II disks known, with characteristic radii of $\sim 400$~au in millimeter dust and $\sim 700$--$900$~au in CO isotopologue gas. Despite its exceptional size and brightness, it lies on the same empirical size--flux relations defined by nearby disk populations, indicating that it represents the high-radius tail of a continuous disk distribution rather than an anomalous outlier.}

\item{The continuum emission reveals a structured hierarchy consisting of an inner core, a pronounced W-shaped double-gap feature at $\sim100$~au, two quasi-regular outer ring--gap pairs at $\sim170$--$230$~au, and a diffuse outer disk skirt. All resolved rings have intrinsic widths of order the local pressure scale height, providing evidence for efficient dust confinement in a gas-rich environment.}

\item {The PV diagrams exhibit the characteristic signature of differential rotation in both $^{12}$CO and $^{13}$CO. No distinct kinematic component associated with an infalling envelope is detected. We quantify the rotation field by fitting a power-law model. The inferred $^{13}$CO slope is close to the Keplerian expectation, yielding a dynamical stellar mass of $M_{\star}=0.88\pm 0.09~M_{\odot}$.}

\item{Using the archival scattered-light image together with the disk orientation derived from the millimeter continuum, we geometrically constrain the height of the scattering-surface as $H_{\rm s}(r)/r \propto (r/1~\mathrm{au})^{1.09\pm0.07}$, consistent with a passively irradiated, flared disk. We then translate this surface shape into the disk flaring angle, $\varphi(r) \propto (r/1~\mathrm{au})^{0.09\pm0.07}$, which implies weak flaring. At the outer edge of the scattered-light disk ($r=204~\mathrm{au}$), we infer $\varphi \sim 0.01$, smaller than the canonical $\varphi \sim 0.05$ typically adopted for classical irradiated disks.}

\item{We derive an observation-anchored temperature structure using the disk flaring angle $\varphi(r)$. The inferred outer disk is cold, with $T_{\rm d}\simeq 6$~K at the outer edge of the directly constrained scattering-surface ($r=204$~au). Over the same radial range ($r=1$--$204$~au), the corresponding area-weighted mean temperature is $\langle T_{\rm d}\rangle=8\pm2$~K.}

\item{The millimeter continuum extends to larger radii than the near-infrared scattered-light, consistent with a vertically settled outer disk and weak vertical mixing. Independent constraints from the observed narrow ring widths, interpreted as limits on turbulent diffusion over their lifetime, place conservative upper bounds on the effective turbulent viscosity, $\alpha \lesssim 10^{-3}$ (and potentially $\lesssim 10^{-4}$), across the outer disk.}

\item {The W-shaped morphology (at $r\sim100$ au) on the dust emission coincides with a scattered-light depression, indicating a perturbation that affects both the disk's surface and midplane. This multilayer signature is consistent with multigap excitation by a single subthermal companion embedded in a low-viscosity disk. Using the observed gap separation and an observation-anchored pressure scale height, we infer a companion mass of $M_{\rm p} = (55 \pm 35)\,M_{\oplus}$.}

\item {The two outer ring-gap pairs (at $r\sim170$--$230$~au) share similar gap widths and depths and show no clear counterpart in the scattered-light image. While we do not exclude a planetary origin, as a distant response to the companion that may drive the W-shaped structure, these properties favor an interpretation in which the dominant perturbation is concentrated toward the midplane. Motivated by this vertical selectivity, we test secular GI as a competing mechanism and find that the disk conditions are compatible with secular GI operating not only at the outermost rings but also across intermediate radii.}

\item {Taken together, V1094~Sco appears to host a hybrid substructure architecture in which weak turbulence plays a central regulatory role. At intermediate radii, a low-mass companion can create multiple gaps through spiral wave excitation, while at larger radii a global midplane instability can organize solids into long-lived rings. In this framework, secular GI does not compete with planet formation; rather, it concentrates solids and establishes disk-wide conditions favorable for subsequent planet growth.}

\end{enumerate}

\begin{acknowledgments}
We thank the anonymous referee for their constructive comments and suggestions, which have improved both the clarity and the content of this manuscript. We also thank T. Nakazato and S. Ikeda for their technical support of the PRIISM imaging. This work was supported by NAOJ ALMA Scientific Research grant Code 2022-22B and JSPS KAKENHI grant (JP26K17220: M.Y., JP26K00741: M.Y. and J.S., JP25K07369: M.N.M., JP26K17198: J.S., JP23K03463: T.M., JP24K07097: T.T., JP25KJ1947: A.S.). H.B.L. and M.T. are supported by the National Science and Technology Council (NSTC) of Taiwan (grant Nos. 113-2112-M-110-022-MY3, 114-2112-M-001-002). J.S. acknowledges support from the KU-DREAM program of Kagoshima University. This paper makes use of the following ALMA data:\\
ADS/JAO.ALMA$\#$2016.1.01239.S,\\
ADS/JAO.ALMA$\#$2017.1.01167.S,\\
ADS/JAO.ALMA$\#$2021.1.00128.L,\\
ADS/JAO.ALMA$\#$2022.1.00875.L.\\
ALMA is a partnership of ESO (representing its member states), NSF (USA) and NINS (Japan), together with NRC (Canada), NSTC and ASIAA (Taiwan), and KASI (Republic of Korea), in cooperation with the Republic of Chile. The Joint ALMA Observatory is operated by ESO, AUI/NRAO and NAOJ.

Data analysis was in part carried out on the multiwavelength Data Analysis System operated by the Astronomy Data Center (ADC) and the Large-scale data analysis system co-operated by the Astronomy Data Center and Subaru Telescope, National Astronomical Observatory of Japan.

This work has made use of data from the European Space Agency (ESA) mission
{\it Gaia} (\url{https://www.cosmos.esa.int/gaia}), processed by the {\it Gaia}
Data Processing and Analysis Consortium (DPAC,
\url{https://www.cosmos.esa.int/web/gaia/dpac/consortium}). Funding for the DPAC
has been provided by national institutions, in particular the institutions
participating in the {\it Gaia} Multilateral Agreement.

The Combined Atlas of Sources with Spitzer IRS Spectra (CASSIS) is a product of the IRS instrument team, supported by NASA and JPL. CASSIS is supported by the ``Programme National de Physique Stellaire'' (PNPS) of CNRS/INSU co-funded by CEA and CNES and through the ``Programme National Physique et Chimie du Milieu Interstellaire'' (PCMI) of CNRS/INSU with INC/INP co-funded by CEA and CNES.

\end{acknowledgments}


\facilities{ALMA, VLT:Melipal, Gaia, Akari, WISE, Herschel, Spitzer, CTIO:2MASS}

\software{
AJISAI (\url{https://github.com/Y-Masayuki/AJISAI}),
AnalysisUtilities \citep{hunter_2023},
Astropy \citep{astropy2022},
CASA \citep{CASATeam2022},
emcee \citep{Foreman2013},
extinction \citep{Barbary2016},
dynesty \citep{speagle_dynesty_2020},
Keplerian Mask Generator (\url{https://github.com/rorihara/Keplerian_Mask_Generator}),
Linmix \citep{kelly_aspects_2007},
matplotlib \citep{Hunter2007},
NumPy \citep{harris_array_2020},
PRIISM \citep{Nakazato2020, Nakazato_priism_2020},
SciPy \citep{virtanen_scipy_2020},
SHIDARE,
SLAM \citep{aso_spectral_2024},
ysoisochrone \citep{deng_ysoisochrone_2025}
}

\restartappendixnumbering 
\appendix
\section{ALMA Data Reduction and CLEAN Imaging Procedure}\label{apx:clean_imaging_details}

\begin{table*}[htbp]
\caption{Summary of ALMA Observations\label{tab:alma_obs}}
{\setlength{\tabcolsep}{3.9pt} 
\begin{tabularx}{\linewidth}{lcccccccc}
\toprule
Project ID & Product & BL Range & Freq & MRS & OST & Flux Calibrator & Obs. Date & Used for \\
           &         & (m)      & (GHz)& (arcsec) & (min) & & (y/m/d) & Analysis \\
(1) & (2) & (3) & (4) & (5) & (6) & (7) & (8) & (9) \\
\midrule
2016.1.01239.S & TM1 & 16--2647 & 225.24 & 2.3  & 2.7  & J1427-4206             & 2017 Jul 7 & cont + CO \\
2017.1.01167.S & TM1 & 92--8548 & 238.75 & 0.8  & 8.1  & J1427-4206             & 2017 Nov 23 & cont only \\
2017.1.01167.S & TM2 & 15--1398 & 238.77 & 3.6  & 4.2  & J1517-2422             & 2018 Jan 17 & cont only \\
2021.1.00128.L & TM1 & 15--2617 & 234.00 & 2.7  & 28.5 & J1427-4206             & 2021 Dec 1 & cont + CO \\
2021.1.00128.L & TM2 & 15--314  & 234.02 & 10.2 & 8.0  & J1517-2422             & 2022 Mar 28 & cont + CO \\
2022.1.00875.L & TM1 & 15--784  & 248.18 & 5.3  & 10.6 & J1924-2914/J1256-0547            & 2023 Jan 12,22 / Mar 4 & cont only \\
\bottomrule
\end{tabularx}
}
\tablecomments{
Column descriptions:
(1) ALMA project ID.
(2) ALMA data product type. TM1 and TM2 denote the long- and short-baseline observations, respectively, within the same project ID.
(3) Range of unprojected baseline lengths (BLs) in meters. We list BL ranges rather than configuration labels because the latter are cycle dependent.
(4) Median observing frequency in GHz.
(5) Maximum recoverable scale (MRS) in arcseconds, given by $0.983 \lambda/D_5$, where $D_5$ is the fifth percentile of $uv$-distance (see the ALMA Technical Handbook).
(6) Total on-source time (OST) in minutes.
(7) Flux calibrator.
(8) Observation date.
(9) Data usage in this work. The CO isotopologue imaging uses three SBs: 2016.1.01239.S (TM1) in \cite{van_terwisga_v1094_2018} and 2021.1.00128.L (TM1 and TM2) in \cite{zhang_alma_2025}. All listed SBs also including 2017.1.101167.S (PI: S. Perez) and 2022.100875.L (PI: I. Cleeves) are used for the Band~6 continuum analysis.}
\end{table*}

We used the Common Astronomy Software Applications package \citep[\texttt{CASA};][]{CASATeam2022} for data calibration and imaging. Each dataset was initially calibrated using the \texttt{CASA} pipeline scripts provided by the ALMA regional centers. Since the observations were conducted over several years, different versions of \texttt{CASA} were used for the initial calibrations. For subsequent processing and imaging, we employed \texttt{CASA} version 6.5. When combining datasets from different epochs, we corrected small astrometric offsets to avoid image blurring and artifacts caused by misalignment. For each dataset, we first produced a continuum image with the CASA {\tt tclean} task to identify the continuum intensity peak and measured its offset from the phase center. We then used {\tt phaseshift} to apply a phase rotation to the visibilities so that the continuum peak is aligned to a common reference direction. Finally, all datasets were assigned a common phase center, (R.A., Dec.)$_\mathrm{ICRS}$ = (16$^\mathrm{h}$08$^\mathrm{m}$36.16$^\mathrm{s}$, $-39\arcdeg$23$'$02.88$''$), using the CASA task {\tt fixplanets}. This task only recognizes J2000 coordinate frame so we temporarily relabeled the FIELD and SOURCE tables from ICRS to J2000 in advance. After applying this task, we restored the original coordinate frame labels in the tables. 

Next, we corrected their flux scales. Specifically, we deprojected the real components of the visibilities and compared them within overlapping $uv$ ranges to derive scaling factors. The task \texttt{gencal} was used to apply the correction to one of the datasets. The shortest baseline dataset of 2021.1.00128.L TM2 (see Table \ref{tab:alma_obs}) is used as the reference for the flux scaling because it offers the most extensive coverage at short $uv$ spacings. The relative flux-scale discrepancies are $5-15\%$ across the datasets. We note that this range is consistent with the expected combination of frequency-dependent intrinsic flux variation between datasets (up to $\sim 13\%$ assuming a millimeter spectral index of $\alpha \sim 2$) and the ALMA Band~6 absolute flux calibration uncertainty of $\sim 5\%$ (ALMA Technical Handbook).

After the data correction, we performed iterative self-calibration on each individual dataset to correct any remaining residual gain errors. In this procedure, a semiautomated self-calibration is applied to generate self-calibrated CLEAN images using our developed tool \texttt{AJISAI}. Details of this framework are described in Appendix~\ref{apx:selfcal}. \texttt{AJISAI} automatically performed the iterative self-calibration to each dataset individually, improving the peak SNR in the image domain by a factor of 1.5–2.0 compared to the non-self-calibrated CLEAN images. After correcting for these gain errors, all datasets were combined using the CASA task \texttt{concat}. Phase-only self-calibration was then performed on the combined dataset using \texttt{AJISAI}, resulting in $\sim$10\% improvement in peak SNR on the image domain, compared with the data-combined one but without the self-calibration procedure. The final CLEAN continuum image from the combined dataset has an observing wavelength of 1.3 mm (232.328 GHz). It also achieves an RMS noise level of $17~\mu\mathrm{Jy~beam}^{-1}$ and a peak intensity of $3975~\mu\mathrm{Jy~beam}^{-1}$, yielding a peak SNR of $\sim230$. The synthesized beam size is $97 \times 74$~mas with a position angle (PA) of $-88\fdg7$, based on the Briggs weighting scheme with a robust parameter of 0.5.

We next assessed the theoretical sensitivity of the combined dataset using the CASA task \texttt{apparentsens}. This task calculates the point source sensitivity for the data by accounting for its imaging parameters such as the weighting scheme and the $uv$ coverage of the data. The resulting theoretical RMS noise was estimated to be $8.2~\mu\mathrm{Jy~beam}^{-1}$ with the same Briggs weighting so that the observed RMS noise compared with the theoretical sensitivity was not greater than a factor of three. The small factor indicates a successful case that the final data product has small calibration errors \citep{bollo_almacal_2024}.

For the $^{12}$CO ($J=2\textrm{--}1$) and $^{13}$CO ($J=2\textrm{--}1$) line measurement sets, we used the subset of datasets that provide velocity channel widths finer than $0.2~\mathrm{km~s^{-1}}$ (Table~\ref{tab:alma_obs}). We applied the same relative flux scaling factors derived from the continuum visibilities (Appendix~\ref{apx:clean_imaging_details}), adopting the short-baseline dataset of 2021.1.00128.L TM2 as the reference, and then applied the continuum self-calibration gain tables to the line data. The flux-scaled line datasets were combined following the same workflow as for the continuum. Continuum subtraction was performed in the $uv$ plane using \texttt{uvcontsub} with \texttt{fitorder}=1 prior to imaging.

\section{AJISAI (Automated Justification-based Imaging and Self-calibration for ALMA Infrastructure)}\label{apx:selfcal}

In the self-calibration procedure, we employed a our newly developed pipeline, Automated Justification-based Imaging and Self-calibration for Alma Infrastructure (\texttt{AJISAI}; /a:ji:sai/)\footnote[6]{\texttt{AJISAI} is publicly available at \url{https://github.com/Y-Masayuki/AJISAI}.}, to perform iterative self-calibration in a fully automated and reproducible manner. This outlines the procedures and parameter choices implemented in the pipeline.

As an initial step, \texttt{AJISAI} determines a reference antenna by querying the \texttt{ANTENNA} table of the measurement set. It retrieves the antenna names and their topocentric positions ($X, Y, Z$). Using the $X$ and $Y$ coordinates, the tool computes the geometric center of the array projected onto the horizontal plane and selects the antenna closest to this center. This choice is intended to minimize baseline-dependent phase errors by selecting an antenna with an average location across the array.

To construct the model for self-calibration, we first perform deconvolution using the \texttt{tclean} task in CASA. We employ multifrequency synthesis imaging with \texttt{nterms = 1} \citep{rau_multi-scale_2011} and the multiscale deconvolution algorithm \citep{cornwell_multiscale_2008}, using scale sizes of $[0,\,1,\,3]\,\theta_{\rm cl}$, where $\theta_{\rm cl}$ denotes the synthesized beam size of the CLEAN image. Briggs weighting with a robustness parameter of 0.5 is adopted throughout. The produced CLEAN model image serves as the initial model for gain calibration.

Gain calibration is carried out using the \texttt{gaincal} task with the following settings. The gain type is set to \texttt{T}, averaging the two parallel-hand polarizations to enhance the SNR. Gain solutions are computed by combining data across scan boundaries (\texttt{combine='scan'}), and four baselines per antenna are used for each solution interval (\texttt{minblperant = 4}). Spectral windows are treated independently. Solutions with SNR $\leq 1.5$ are discarded, but the associated visibilities are retained unflagged by using \texttt{applymode='calonly'} in the \texttt{applycal} step.

We perform three rounds of phase-only self-calibration followed by a single round of amplitude self-calibration. During the phase calibration stages, the solution intervals are progressively shortened from ``inf'' (one solution per scan) to $6\times$ the integration time and finally to $3\times$ the integration time, where a single integration typically corresponds to a few seconds in the ALMA observations. The amplitude self-calibration is performed with a solution interval of ``inf''. The CLEAN image with the highest peak SNR in the image domain among all iterations is adopted as the final self-calibrated image.

\section{Performance of PRIISM Imaging}\label{apx:priism_image}

In this section, we describe the PRIISM imaging procedure and assess the performance of the reconstructed images through a set of quantitative tests. Section~\ref{apx:priism_imaging} summarizes the PRIISM imaging procedure and the selection of regularization parameters. In Section~\ref{apx:effective_resolution}, we estimate the effective spatial resolution of the PRIISM model image using a point-source injection method. Section~\ref{apx:restored_image} describes the construction of the PRIISM restored image used for noise characterization. In Section~\ref{apx:goodnessfit}, we evaluate how well the PRIISM model image reproduces the observed visibilities in the visibility domain using reduced chi-square statistics.
\subsection{PRIISM Imaging Procedure}\label{apx:priism_imaging}

We describe the procedure of super-resolution imaging with $\tt PRIISM$ \citep[version 0.11.5;][]{Nakazato2020, Nakazato_priism_2020}. This approach produces smoother model continuum images and achieves improved effective spatial resolution compared to CLEAN, owing to its regularized maximum-likelihood optimization with $\ell _1$+TSV imaging and the CV scheme as illustrated in \citet{yamaguchi_super-resolution_2020}. The model image is reproduced by minimizing a cost function, which is the chi-square error between the visibility model derived by the model image and observed visibility, accompanied by two regularization terms, namely $\ell_1$-norm and TSV, which are parameterized by the parameters $\Lambda_{l}$ and $\Lambda_{tsv}$, respectively. This approach has been widely applied to circumstellar disks observed by ALMA to date \citep{yamaguchi_super-resolution_2020, yamaguchi_alma_2021, Yamaguchi_alma_2025ApJ, chou_probing_2025, shoshi2025b, shoshi_ring-gap_2026}. Its ability to produce high image fidelity has been demonstrated using ALMA data \citep{yamaguchi_alma_2024, shoshi2025b}. 

Using the self-calibrated visibility data, we reconstructed a set of PRIISM model images over a grid of $(\Lambda_{l}, \Lambda_{tsv})$ values. We explore $(\Lambda_{l}, \Lambda_{tsv})$ over the grid $[(10^{3}, 10^{4}, 10^{5}, 10^{6}, 10^{7}), (10^{10}, 10^{11}, 10^{12}, 10^{13}, 10^{14})]$. We then selected the optimal pair of $(\Lambda_{l}, \Lambda_{tsv})$ using 10-fold CV, adopting the model that minimizes the cross-validation error (CVE) \citep{yamaguchi_super-resolution_2020, yamaguchi_alma_2024}. The final pairs for $(\Lambda_{l}, \Lambda_{tsv})$ with the minimum CVE provided ($10^{5}$, $10^{12}$).

\subsection{Effective Spatial Resolution}\label{apx:effective_resolution}

Because PRIISM reconstructs a model image without beam convolution, we measure the effective spatial resolution $\theta_{\rm eff}$ using the point-source injection method described in \citet{yamaguchi_alma_2021}. We inject an artificial point source into the observed visibility data, adopting a flux density of $5\%$ of the total continuum flux of the target. The source is placed in an emission-free region and at a location where the recovered image-plane flux remains consistent with the injected value \citep{shoshi_ring-gap_2026}.

We repeat the injection in four directions (east, west, north, and south) and perform independent PRIISM reconstructions for each case using the same regularization parameters as in the optimal reconstruction of the original dataset. In the reconstructed images, the injected sources appear as compact, approximately Gaussian components. We fit each component with an elliptical Gaussian function and define the fitted FWHM as the effective spatial resolution.

We measure an averaged effective spatial resolution of $\theta_{\rm eff} = 40 \times 30$~mas at a PA of $-77\arcdeg$. The variation among the four injection directions is within a few percent, and we therefore adopt the average as the representative value. This definition quantifies the empirical width of the reconstructed point-source response and provides a practical measure of the smallest recoverable structure scale under the adopted imaging conditions.

\subsection{Restored Image Construction}\label{apx:restored_image}

As described in Section~\ref{apx:effective_resolution}, PRIISM imaging produces a model image without beam convolution. Consequently, the pixel values are expressed in units of $\rm Jy~pixel^{-1}$ and do not directly correspond to a noise estimate in $\rm Jy~beam^{-1}$. For analyses that require a beam-convolved image with an empirical rms noise estimate, we therefore constructed a restored image following the procedure described in \citet{Yamaguchi_alma_2025ApJ}.

The restored image is defined as $I_r = I_m * B + I_d$, where $I_m$ is the PRIISM model image, $B$ is an elliptical Gaussian beam approximating the main lobe of the synthesized beam, and $I_d$ is the residual map obtained as the inverse Fourier transform of the residual visibilities between the model and observed visibilities. This restoration procedure is conceptually similar to the standard CLEAN restoration \citep{thompson_interferometry_2017}.

For the present dataset, we constructed the restored image using Briggs robust $=2.0$ and a $uv$ taper of $500~\mathrm{k}\lambda$ in order to enhance sensitivity to extended emission. The resulting synthesized beam is $362 \times 331$~mas (PA $=87\arcdeg$). The rms noise level was estimated from emission-free regions of the restored image and is $23~\mu\mathrm{Jy~beam^{-1}}$.

The restored image is used for estimating the rms noise level and visualizing faint extended emission, while the morphological analysis is based on the PRIISM model image.

\subsection{Assessment of Image Fidelity and Goodness of Fit}\label{apx:goodnessfit}

\begin{figure}[ht]
\centering
\includegraphics[width=0.46\textwidth]{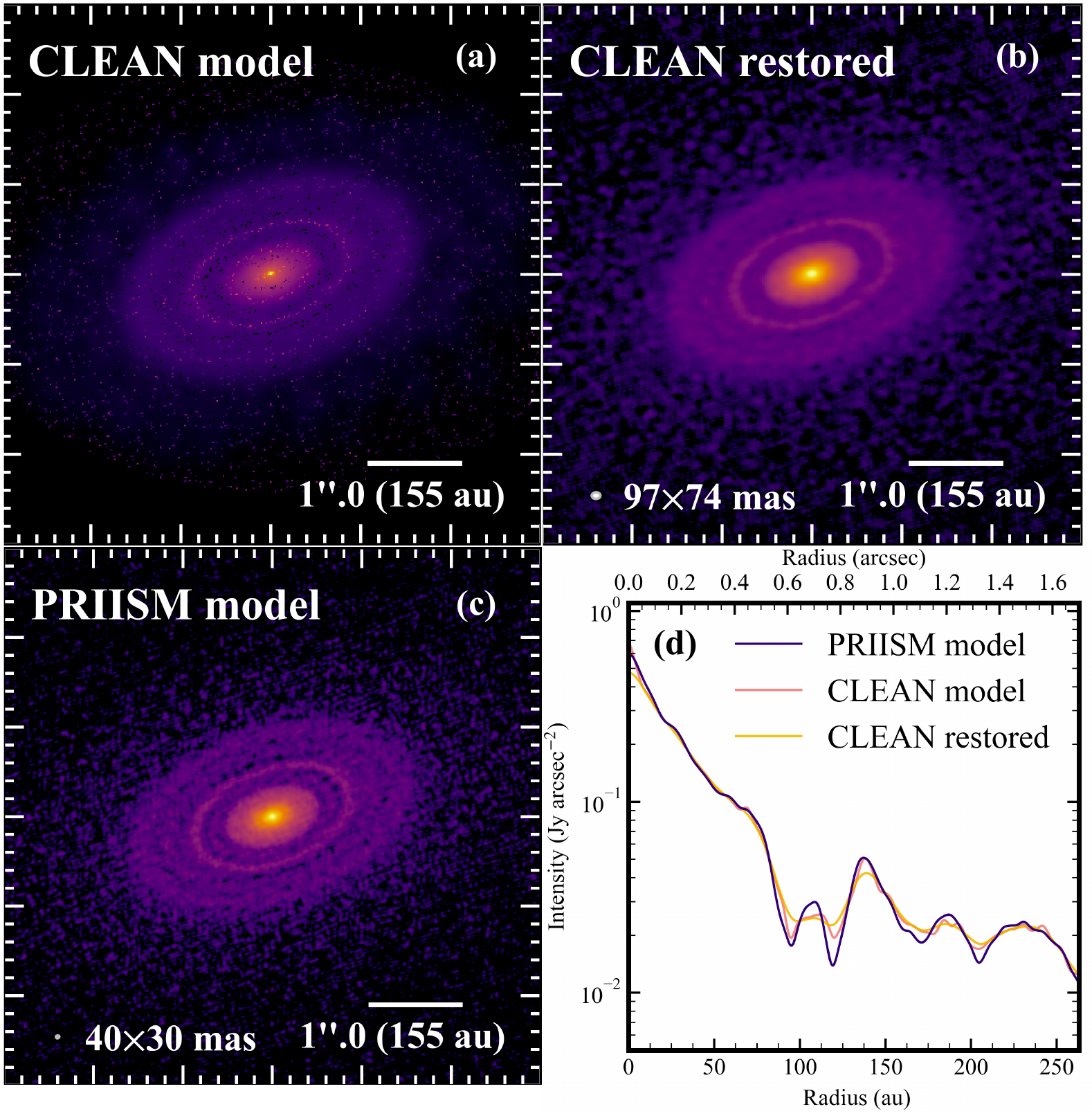}
\caption{Gallery of ALMA continuum images reconstructed by different imaging techniques, such as (a) CLEAN model, (b) restored CLEAN image, and (c) PRIISM model. (d) Azimuthally average radial intensity profiles for (a)-(c).}
\label{fig:images_comparison}
\end{figure}

\begin{figure*}[t]
\centering
\begin{minipage}{0.98\textwidth}
    \centering
    \includegraphics[width=\textwidth]{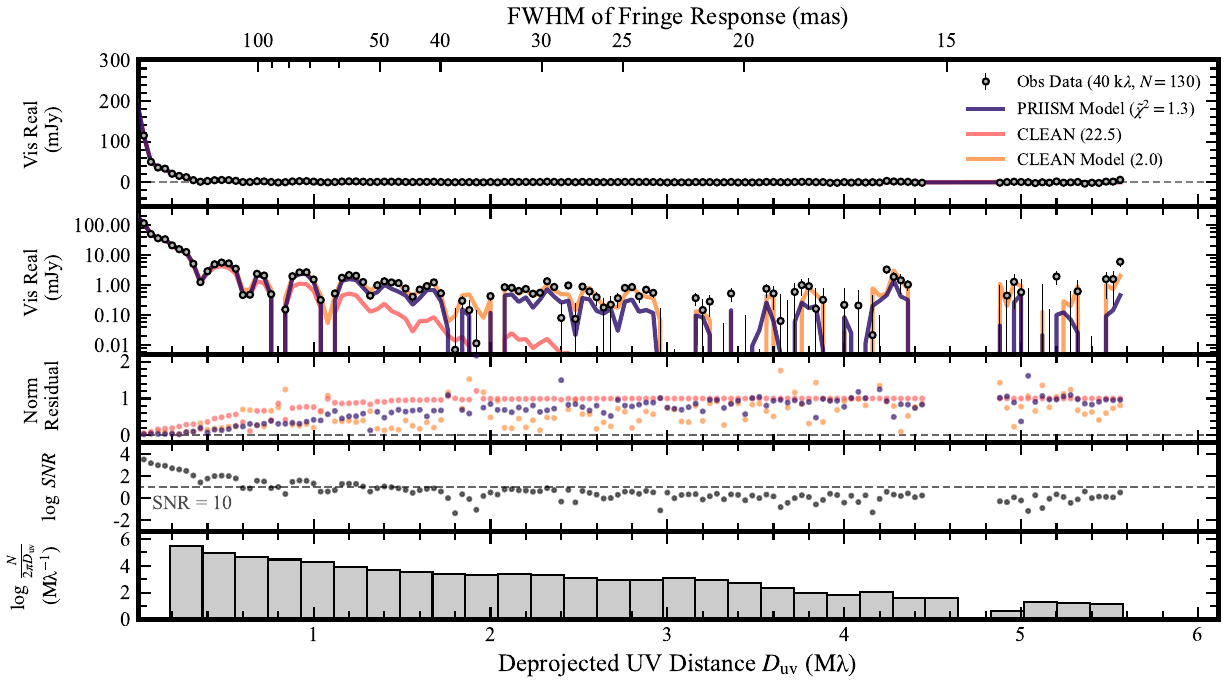}
    \captionof{figure}{
        The azimuthally averaged radial visibility profile of the V1094 Sco disk. The observed visibility data are shown by dots, and the visibility models with PRIISM model, beam-convolved CLEAN image, and CLEAN model are represented by black, red, and orange lines, respectively. The data are binned every $40k\lambda$. The reduced $\chi^2$ values calculated from the observed data and the models are shown in the labels of the top panels. The left-hand panels of each target display, from top to bottom, the amplitude of the real part of the visibility, its logarithmic scale, the normalized residual between the observation and the model, the SNR of visibility within each bin, and the data density of each bin in $uv$ space. The SNR is the ratio of the real part amplitude to its noise. The details are described in an appendix in
        \citet{yamaguchi_alma_2024}.
    }
    \label{fig:vis_profile}
\end{minipage}
\end{figure*}

Figure~\ref{fig:images_comparison} provides a direct comparison between the PRIISM and CLEAN reconstructions. The CLEAN model image shows patchy structures, which can arise from the behavior of the multiscale CLEAN algorithm when approximating smoothly varying emission. Such representations are known to reproduce extended astrophysical structures only imperfectly and can introduce artificial small-scale fluctuations, particularly for disks with gradual radial and azimuthal brightness variations \citep{cornwell_multiscale_2008}.

In contrast, PRIISM employs a regularization scheme based on the TSV functional, which favors spatially smooth but data-consistent solutions and suppresses unphysical small-scale fluctuations. As a result, PRIISM yields surface-brightness distributions that more closely follow the expected smooth disk morphology.

To quantify how well each reconstruction reproduces the observations, we compute reduced chi-square values, $\tilde{\chi}^{2}$, in the visibility domain using azimuthally averaged visibility profiles (see Appendix~4 of \citealt{yamaguchi_alma_2024} for the definition). As shown in Figure~\ref{fig:vis_profile}, the PRIISM model achieves $\tilde{\chi}^{2} \approx 1$, indicating a match to the data, whereas the CLEAN model yields a larger value.

Taken together, our tests demonstrate that PRIISM provides a higher-fidelity reconstruction of the continuum emission than CLEAN for this dataset. For this reason, and because our scientific interpretation relies on resolving smooth substructures with high effective spatial resolution, we adopt the PRIISM reconstruction as the primary basis for the continuum analyses presented in this work.

\section{SHIDARE (SED-based Holistic Integrated Derivation of All Stellar Radiative Parameters)}\label{apx:stellar_properties}

\begin{figure}[t]
\centering
\begin{minipage}{0.48\textwidth}
    \centering
    \includegraphics[width=\linewidth]{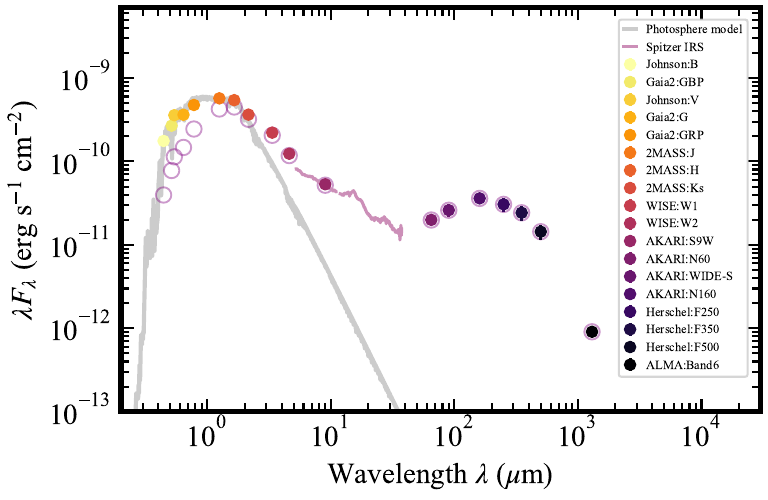}
    \caption{SED of V1094 Sco. The SED has been dereddened using the best-fit $A_V$ value listed in Table~\ref{tab:stellar_property}. The gray solid curve shows the best-fit stellar photosphere model (BT-Settl model shown here), and the purple solid curve represents the \textit{Spitzer}/IRS spectrum. The scattered symbols indicate photometric measurements, with different colors corresponding to the telescope names shown in the right-hand legend. The original (reddened) data points are also plotted as purple circles for reference.}
    \label{fig:sed}
\end{minipage}
\hfill
\begin{minipage}{0.48\textwidth}
    \centering
    \includegraphics[width=\linewidth]{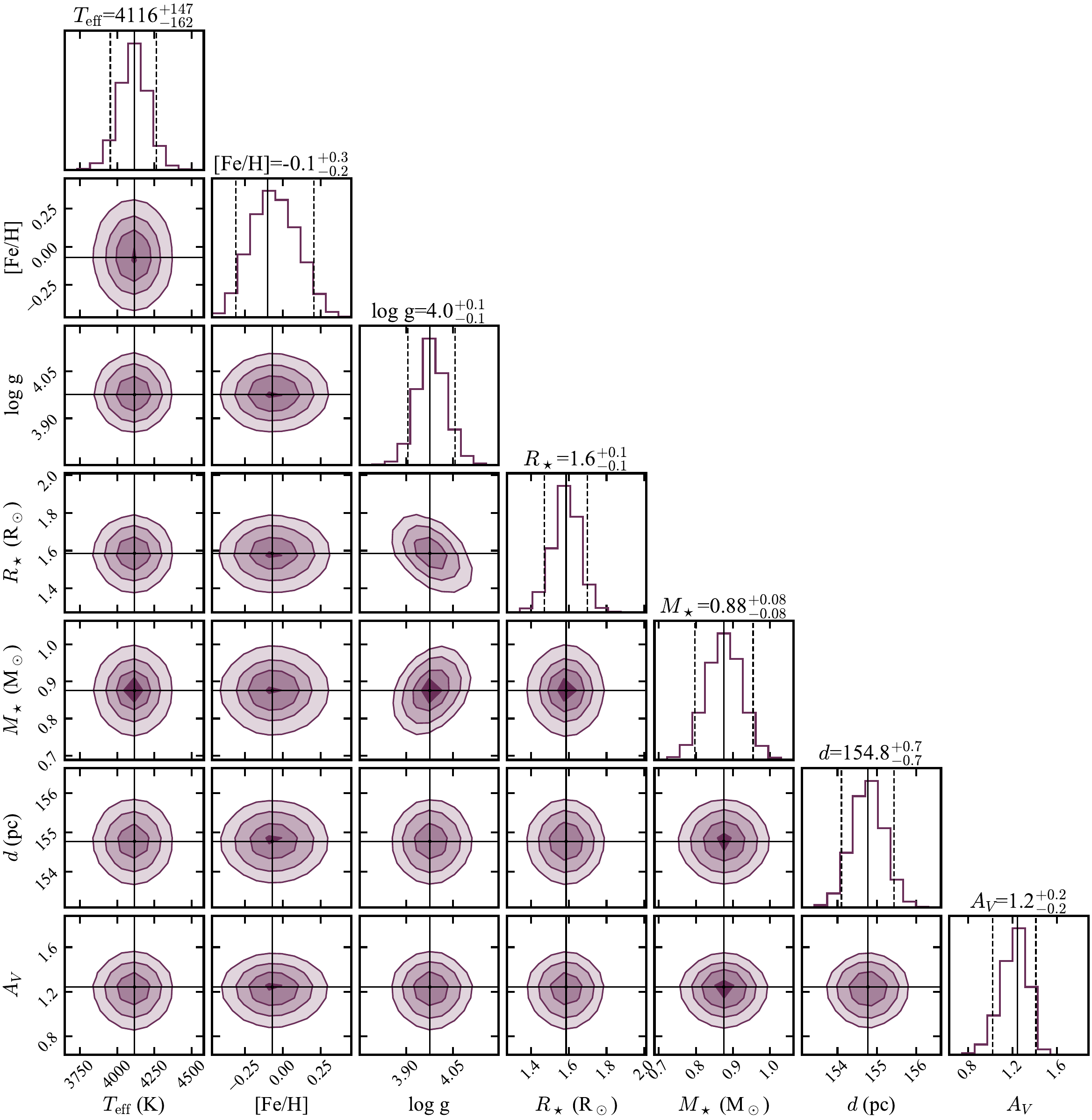}
    \caption{Posterior corner plot of the stellar parameters derived from the Bayesian SED fitting with \texttt{SHIDARE}. The two-dimensional contours represent the highest density intervals enclosing 68\%, 95\%, and 99.7\% of the total posterior probability, corresponding to the 1$\sigma$, 2$\sigma$, and 3$\sigma$ credible regions. The diagonal panels show the marginalized one-dimensional posterior distributions for each parameter.}
    \label{fig:sed_corner}
\end{minipage}
\end{figure}

\begin{table}[t!]
\centering
\caption{Adopted priors and posterior stellar properties derived from the \texttt{SHIDARE} Bayesian SED fitting.}
\label{tab:stellar_property}
\renewcommand{\arraystretch}{1.3}
\begin{tabularx}{\columnwidth}{l@{\hspace{1pt}}Y@{\hspace{55pt}}Y@{\hspace{40pt}}r@{\hspace{-50pt}}}\hline\hline
Property & Prior Distribution & Posterior Value & Prior Information\\
\hline
$T_{\rm eff}$ (K) & $\mathcal{U}(3500,\,5000)$ & $4116^{+147}_{-162}$ & typical TTS range \\
$[{\rm Fe/H}]$ (dex) & $\mathcal{U}(-0.5,\,0.5)$ & $-0.07^{+0.28}_{-0.24}$ & typical TTS range \\
$R_\star~(R_{\odot})$ & $\mathcal{U}(0.5,\,2.5)$ & $1.59^{+0.11}_{-0.12}$ & typical TTS range \\
$d$ (pc) & $\mathcal{N}(154.8,\,0.8)$ & $154.8^{+0.7}_{-0.7}$ & \textit{Gaia} DR3$^{a}$ \\
$A_V$ (mag) & $\mathcal{N}(1.7,\,0.5)$ & $1.24^{+0.17}_{-0.22}$ & VLT/X-shooter$^{b}$ \\
$M_\star~(M_{\odot})$ & $\mathcal{N}(0.88,\,0.09)$ & $0.88^{+0.07}_{-0.07}$ & ALMA (this work) \\
\hline
$\log g$ (cm~s$^{-2}$)         & (derived) & $3.98^{+0.07}_{-0.07}$  & $\cdots$ \\
$L_\star~(L_{\odot})$  & (derived) & $0.64^{+0.15}_{-0.13}$   & $\cdots$ \\
$T_{\mathrm{bol}}$ (K) & (derived) & $2867^{+18}_{-18}$       & $\cdots$ \\
$L_{\mathrm{bol}}~(L_\odot)$ & (derived) & $0.80^{+0.01}_{-0.01}$ & $\cdots$ \\
Age (Myr)   & (derived) & $2.4^{+1.7}_{-1.0}$      & $\cdots$ \\
\hline
\end{tabularx}
\vspace{4pt}
\noindent\makebox[\columnwidth][c]{%
\parbox{0.98\columnwidth}{%
\footnotesize
\textbf{Notes.} $\mathcal{U}(a,b)$ denotes a uniform prior between $a$ and $b$, and $\mathcal{N}(\mu,\sigma)$ denotes a normal prior with mean $\mu$ and standard deviation $\sigma$. Uncertainties represent the 68\% highest density interval of the posterior distributions. The maximum a posteriori estimate was adopted as the representative value.
\tablenotetext{a}{\textit{Gaia} DR3 parallax \citep{Gaia2023}.}
\tablenotetext{b}{VLT/X-shooter spectroscopy \citep{alcala_x-shooter_2017}.}
}%
}
\end{table}

With the advent of \textit{Gaia} and ALMA, which respectively provide precise stellar distances and dynamical stellar mass measurements, stellar parameters that were previously inferred from conventional pre-\textit{Gaia} SED fitting \citep{tsukagoshi_detection_2011} can now be determined with much higher accuracy. In this work, we refine the stellar properties of V1094~Sco using our newly developed tool, Sed based Holistic Integrated Derivation of All stellar Radiative paramEters (\texttt{SHIDARE}; /shi:da:re/), a Bayesian SED analysis pipeline designed to automatically collect broad wavelength radiative information and to infer stellar parameters in a self-consistent and integrated manner.

\texttt{SHIDARE} compiles photometric fluxes through the \texttt{astroquery} interface \citep{ginsburg_astroquery_2019}, incorporating APASS DR9 \citep[Johnson $B$ and $V$;][]{Henden2014}, \textit{Gaia} \citep{gaia_collaboration_gaia_2018,Gaia2023}, Two Micron All Sky Survey \citep[2MASS;][]{Cutri2003}, AKARI \citep{Yamamura2010}, Wide-field Infrared Survey Explorer \citep[WISE;][]{Cutri2014}, and \textit{Herschel} \citep{pilbratt_herschel_2010,griffin_herschel_2010}. The pipeline also ingests the ALMA fluxes used in this work and the mid-infrared spectra retrieved from the \textit{Spitzer}/IRS archive through CASSIS \citep{lebouteiller_cassis_2011}. In practice, the APASS DR9, \textit{Gaia}, and 2MASS bands primarily constrain the stellar photospheric parameters, whereas the longer wavelength data are used mainly to derive the bolometric temperature ($T_{\mathrm{bol}}$) and luminosity ($L_{\mathrm{bol}}$).

\texttt{SHIDARE} models the observed SED as
\begin{equation}
F_\nu^{\mathrm{obs}}(\lambda)
= \left( \frac{R_\star}{d} \right)^{2}
  F_\nu^{\mathrm{mod}}(T_{\mathrm{eff}}, \log g, [{\rm Fe/H}], \lambda)
  \, 10^{-0.4\,A_\lambda},
\label{eq:sed_model}
\end{equation}
where $F_\nu^{\mathrm{mod}}$ denotes the synthetic stellar flux density computed from atmosphere models, $(R_\star/d)^2$ scales the intrinsic flux to the observed level, and $A_\lambda$ accounts for wavelength-dependent interstellar attenuation.

The SED fitting is carried out within a Bayesian framework inspired by the concept of Bayesian model averaging \citep{vines_dense_2022}. We employ the nested sampling algorithm implemented in the \texttt{dynesty} package \citep{speagle_dynesty_2020}, which computes the Bayesian evidence for each model while sampling the posterior distributions of the stellar parameters. We consider four widely used stellar atmosphere model libraries for T Tauri stars, namely BT Settl, BT NextGen, BT Cond \citep{allard_models_2012,allard_bt-settl_2013}, and PHOENIX v2 \citep{husser_new_2013}. Independent fits are performed for each model grid, and the resulting posterior distributions are combined using Bayesian model averaging with weights proportional to the corresponding Bayesian evidence. This procedure mitigates systematic biases associated with any single stellar atmosphere model grid.

A distinctive feature of \texttt{SHIDARE} is that the stellar surface gravity is not treated as a free parameter. Instead, the pipeline enforces physical consistency by deriving $\log g$ from the independently constrained stellar mass and the SED-derived radius through $g = G M_\star / R_\star^2$. Here, $M_\star$ is anchored to the $^{13}$CO Keplerian rotation measured with ALMA (Section~\ref{sec:rotation_curve}). This design reduces the degeneracy between $g$ and $R_\star$ in broadband SED fitting while directly incorporating external physical information into the inference.

During each grid fit, \texttt{SHIDARE} linearly interpolates synthetic spectra in $T_{\mathrm{eff}}$--$\log g$--[Fe/H] space, scales the model fluxes by $(R_\star/d)^2$, and evaluates wavelength-dependent extinction using the \citet{fitzpatrick_correcting_1999} law as implemented in \texttt{extinction} \citep{Barbary2016}, adopting a total to selective extinction ratio of $R_V = 3.1$ that is characteristic of the diffuse interstellar medium \citep{fitzpatrick_analysis_2019}. The visual extinction $A_V = 1.7 \pm 0.5$ measured from VLT/X-shooter spectroscopy \citep{alcala_x-shooter_2017} is adopted as prior information.

The adopted priors and posterior estimates are summarized in Table~\ref{tab:stellar_property}. The observed and best-fit SEDs are shown in Figure~\ref{fig:sed}, and the posterior distributions are shown in Figure~\ref{fig:sed_corner}. The resulting parameters are consistent with those of pre-main-sequence, low-mass stars \citep{yamashita_measurements_2020,flores_effects_2022}, supporting the classification of V1094~Sco as a young T~Tauri star. The metallicity $[\mathrm{Fe/H}]$ remains weakly constrained because broadband photometry is largely insensitive to the narrow absorption features that carry most of the metallicity information.

The \texttt{SHIDARE} fitting directly constrains the stellar effective temperature $T_{\rm eff}$, while the stellar luminosity $L_\star$ is derived from the fitted stellar radius and effective temperature as $L_\star = 4\pi R_\star^2 \sigma_{\rm SB} T_{\rm eff}^4$, where $\sigma_{\rm SB}$ is the Stefan--Boltzmann constant. Because the position of a source in the Hertzsprung--Russell diagram is primarily set by $(T_{\rm eff},\,L_\star)$, we estimate the stellar age by comparing the inferred $(T_{\rm eff},\,L_\star)$ with stellar evolutionary models. This step is performed with the magnetic evolutionary tracks of \citet{feiden_magnetic_2016} using \texttt{ysoisochrone} \citep{deng_ysoisochrone_2025}, which returns posterior distributions for the stellar age and mass. Using the inferred $T_{\rm eff}$ and $L_\star$, we obtain an age of $2.4^{+1.7}_{-1.0}$ Myr and an evolutionary-track mass of $M_{\star}=0.8^{+0.2}_{-0.1}~M_\odot$, consistent with the ALMA dynamical mass of $0.88\pm0.09~M_\odot$.

For completeness, \texttt{SHIDARE} also computes $T_{\mathrm{bol}}$ and $L_{\mathrm{bol}}$ from the best-fit dereddened SED using the standard definitions. The pipeline performs two independent numerical integrations: trapezoidal integration in frequency space and cubic interpolation with \texttt{scipy.interpolate.PchipInterpolator} \citep{virtanen_scipy_2020}. The two estimates agree within a few percent; their mean values are adopted as the final results and half their difference is taken as an uncertainty. This procedure yields $T_{\mathrm{bol}} = 2867 \pm 18~\mathrm{K}$ and $L_{\mathrm{bol}} = 0.80 \pm 0.01~L_\odot$, consistent with a Class~II object whose SED is dominated by stellar photospheric emission with a moderate disk contribution \citep{Chen1995}.

\section{Azimuthal Selection Bias in CO Gas Disk Radii Measured from Deprojected Moment Zero Maps}\label{apx:gasdisksize_bias}

\begin{figure}[h!]\label{fig:azimuthal_selection_COdiskradii}
\centering
\includegraphics[width=0.46\textwidth]{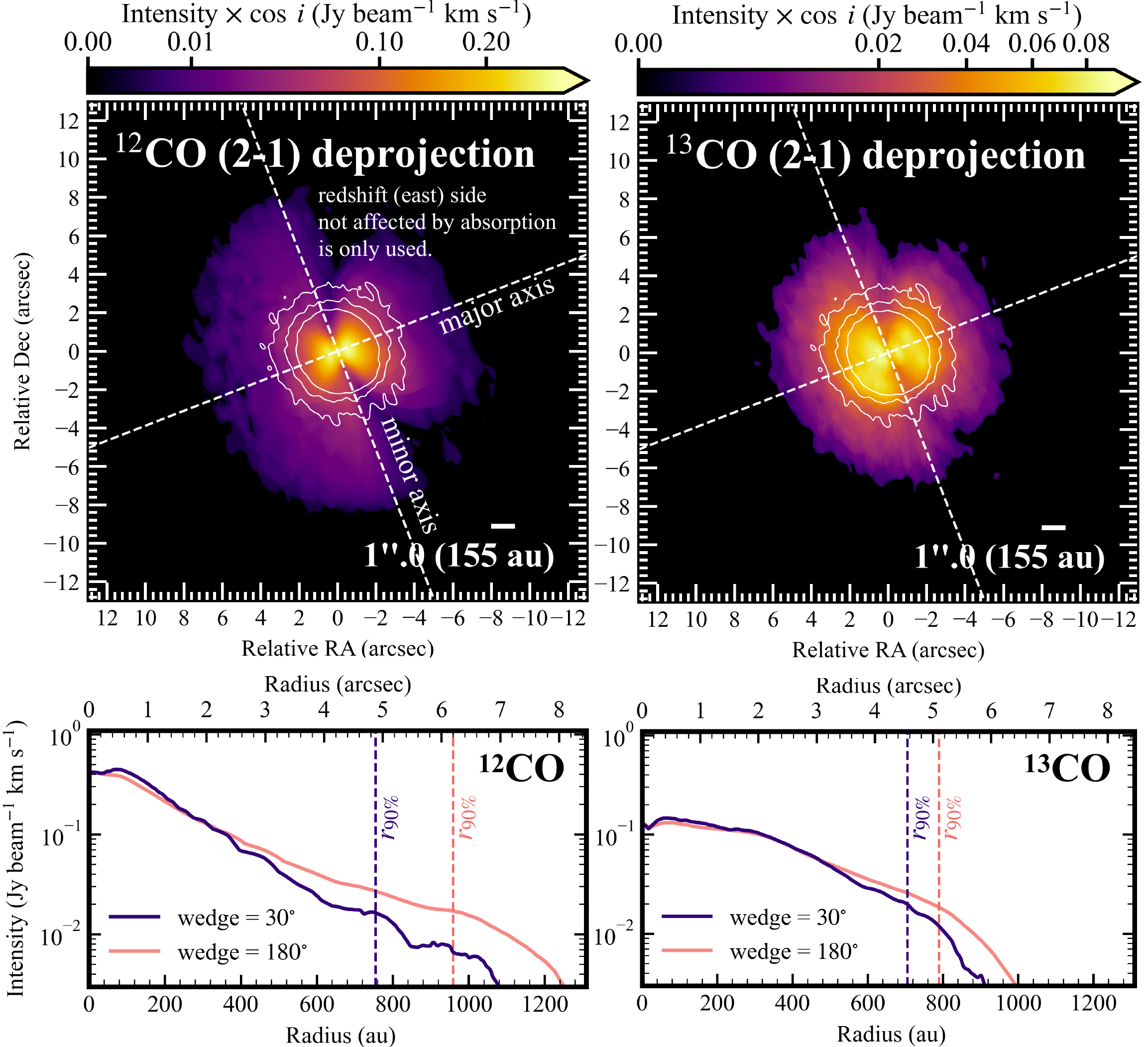}
\caption{Top: Deprojected $^{12}$CO (left) and $^{13}$CO (right) moment zero maps of the V1094 Sco disk. Bottom: Azimuthally averaged radial intensity profiles extracted from the deprojected $^{12}$CO (left) and $^{13}$CO (right) images using different azimuthal wedges. The profiles are computed using only the redshifted (eastern) side of the disk, where foreground absorption is negligible. The purple and red curves show profiles derived from wedges with full opening angles of $30\arcdeg$ and $180\arcdeg$, respectively, centered on the disk's major axis.}
\end{figure}

Measuring gas disk radii from CO moment zero maps is commonly based on a geometric deprojection using the disk inclination and position angle derived from continuum emission, followed by a curve-of-growth analysis \citep{ansdell_alma_2018, law_molecules_2021-1,long_gas_2022,deng_alma_2025,zallio_12_2026, agurto-gangas_alma_2025, ruiz-rodriguez_alma_2025}. This procedure implicitly assumes that the emitting surface can be mapped back to the disk plane through a purely two-dimensional transformation, effectively treating the disk as geometrically flat. However, CO emission, particularly from optically thick $^{12}\mathrm{CO}$, predominantly traces elevated layers above the midplane. For an inclined disk, the projected minor axis coordinate contains an additional contribution from the vertical height of the emitting surface, such that the sky plane coordinate can be expressed as $y_{\mathrm{sky}} = y_{\mathrm{disk}}\cos i + z(r)\sin i$, where $y_{\mathrm{sky}}$ is the observed coordinate along the disk minor axis in the plane of the sky, $y_{\mathrm{disk}}$ is the corresponding coordinate in the disk midplane, and $z(r)$ denotes the height of the CO emitting surface above the midplane. Applying a flat deprojection ($z=0$) therefore introduces a systematic offset $y_{\mathrm{disk}} \rightarrow y_{\mathrm{disk}} + z(r)\tan i$, which artificially stretches the emission along the minor axis. As a result, gas disk radii derived from deprojected moment zero maps can be biased to larger values, with the magnitude of the effect increasing for higher emitting surfaces and larger inclinations.

To quantify this effect in V1094 Sco, we performed a controlled test in which the radius measurement was repeated while varying the azimuthal range included in the curve-of-growth analysis. The radius measurement follows exactly the same procedure as in Section~\ref{sec:results_disksize}, that is, a standard deprojection using the continuum-based inclination and position angle, restriction to the redshifted (eastern) side of the disk where foreground absorption is negligible, and a curve-of-growth analysis applied to the deprojected moment zero maps; the only difference from the main analysis is that we vary the azimuthal opening angle $\Delta\phi$. From the resulting moment zero maps, we retained only the emission within a wedge centered on the disk major axis with a full opening angle $\Delta\phi~(\in[0\arcdeg,180\arcdeg])$, and measured the radius enclosing $90\%$ of the total flux for each $\Delta\phi$.

Figure~\ref{fig:azimuthal_selection_COdiskradii} shows that the derived gas disk radius exhibits a clear and monotonic dependence on the adopted azimuthal range. For ${}^{12}\mathrm{CO}(2-1)$, the radius decreases from $977 \pm 38~\mathrm{au}$ when using a full azimuthal average ($\Delta\phi = 180\arcdeg$) to $762 \pm 38~\mathrm{au}$ for a major axis centered wedge with $\Delta\phi = 30\arcdeg$, corresponding to a reduction of $\simeq 22\%$. For ${}^{13}\mathrm{CO}(2-1)$, the same procedure yields a smaller but still systematic decrease consistent with its deeper emitting layer, from $790 \pm 39~\mathrm{au}$ to $707 \pm 39~\mathrm{au}$ ($\simeq 11\%$). These results demonstrate that adopting a near full azimuthal average overweights minor axis directions, where vertical projection effects are maximized, and can therefore lead to a systematic overestimate of CO gas disk radii compared to measurements anchored to the major axis.

In this work, we employ the gas disk radius measured within a major axis centered wedge with the small opening angle of $\Delta\phi = 30\arcdeg$ as a representative value. This choice is motivated by the result that narrower azimuthal selections systematically reduce the influence of minor axis projection effects, while still retaining sufficient SNR for a robust curve-of-growth analysis. We emphasize that $\Delta\phi = 30\arcdeg$ should not be regarded as a unique or formally optimal choice, but rather as a conservative and reproducible compromise between minimizing geometric bias associated with vertically extended CO emitting layers and preserving statistical stability in the radius determination.

\section{Identification and Quantification of Dust Disk Substructures}
\label{apx:substructure_definition}

We provide the formal definitions and methodological details underlying the identification and characterization of dust disk substructures presented in Section~\ref{sec:results_substructures}. These definitions are included here to ensure reproducibility while maintaining a streamlined presentation in the main results.

\subsection{Radial Intensity Profile}

All substructure measurements are based on the azimuthally averaged radial intensity profile, $I_{\nu}(r)$, extracted from the deprojected dust continuum images. The averaging is performed within a wedge centered on the disk semi-major axis, excluding regions near the minor axis where projection effects and beam smearing degrade the fidelity of radial features.

\subsection{Derivative--based Identification of Substructures}

To identify local morphological features in the radial intensity profile, we adopt a derivative-based approach following \citet{yamaguchi_alma_2024}. We define the normalized radial derivative as
\begin{equation}
\mathcal{D}(r) \equiv \frac{1}{I_{\nu}(r)} \frac{d I_{\nu}(r)}{dr}.
\end{equation}

Local extrema and curvature changes in $\mathcal{D}(r)$ are used to classify distinct types of substructures:

\begin{itemize}
\item Rings (B): Local maxima in $I_{\nu}(r)$, identified at locations where $\mathcal{D}(r)=0$ and $d\mathcal{D}/dr < 0$.

\item Gaps (D): Local minima in $I_{\nu}(r)$, identified where $\mathcal{D}(r)=0$ and $d\mathcal{D}/dr > 0$.

\item Inflection points (I): Concave features embedded within an overall monotonic decline, where $\mathcal{D}(r) < 0$ and $d\mathcal{D}/dr > 0$, corresponding to a change in curvature rather than a full ring--gap pair.
\end{itemize}

\subsection{Definition of Inflection Radius}

For inflection features, the characteristic radius $r_{\rm inf}$ is defined as the location that maximizes the deviation of the observed profile from a straight line connecting the adjacent extrema in $\mathcal{D}(r)$. Specifically,
\begin{equation}
r_{\rm inf}
= \argmax_{r} \left[ \ell(r) - I_{\nu}(r) \right],
\end{equation}
subject to the condition $\ell(r) \ge I_{\nu}(r)$, where $\ell(r)$ denotes the linear interpolation between the neighboring minimum and maximum of $\mathcal{D}(r)$. These inflection features are used to identify broad transitions such as shoulder and disk skirt.

\subsection{Gap Width and Depth}

For each ring--gap pair, we define a characteristic edge intensity as
\begin{equation}
I_{\rm edge} = \frac{1}{2} \left[ I_{\nu}(r_{\rm ring}) + I_{\nu}(r_{\rm gap}) \right].
\end{equation}

The inner and outer edges of the gap, $(r_{\rm in}, r_{\rm out})$, are defined as the radii where $I_{\nu}(r) = I_{\rm edge}$. The absolute gap width and normalized width are then given by
\begin{equation}
\Delta_{\rm I,unit} = r_{\rm out} - r_{\rm in},
\qquad
\Delta_{\rm I} = \frac{r_{\rm out} - r_{\rm in}}{r_{\rm out}}.
\end{equation}

The gap depth is defined as the intensity contrast between the ring and the gap minimum,
\begin{equation}
\delta_{\rm I} = \frac{I_{\nu}(r_{\rm ring})}{I_{\nu}(r_{\rm gap})}.
\end{equation}

\bibliographystyle{aasjournalv7}
\bibliography{references,additional_references}

\end{document}